\documentclass[12pt]{iopart}

\usepackage[utf8x]{inputenc}
\usepackage{iopams}  

\usepackage{mathrsfs}
\usepackage{slashed}
\usepackage{graphicx}
\usepackage{dsfont}
\usepackage{longtable}
\usepackage{tcolorbox}
\usepackage{xcolor}
\definecolor{lcolor}{rgb}{0.5,0,0}
\definecolor{citcolor}{rgb}{0,0.3,0.0}

\newlength{\mycol}
\usepackage[breaklinks,colorlinks,citecolor=citcolor,urlcolor=blue,linkcolor=lcolor]{hyperref}

\newcommand{\trento}{{\rm T}$\mathrel{\protect\raisebox{-2.1pt}{{\rm R}}}${\rm ENTo}}

\begin{document}

\title{The smallest fluid on earth}

\author{Bj\"orn Schenke}
\address{Physics Department, Brookhaven National Laboratory\\ Bldg. 510A, Upton, NY 11973, USA}
\ead{bschenke@bnl.gov}
\vspace{10pt}
\begin{indented}
\item[]May 2021
\end{indented}

\begin{abstract}
    High energy heavy ion collisions create quark gluon plasmas that behave like almost perfect fluids. Very similar features to those that led to this insight have also been observed in experimental data from collisions of small systems, involving protons or other light nuclei. We describe recent developments aimed at understanding whether, and if so how, systems that produce relatively few particles (orders of magnitude less than in typical heavy ion collisions) and are only one to a few times the size of a proton, can behave like fluids. This involves a deeper understanding of fluid dynamics and its applicability, improvements of our understanding of the initial geometry of the collisions by considering fluctuations of the proton shape, as well as advancements in the calculation of initial state effects within an effective theory of quantum chromodynamics, which can affect the observables that are used to study fluid behavior. We further address open questions and discuss future directions. 
\end{abstract}

\section{Introduction}

Matter produced in collisions of heavy ions at high energy as performed at the Relativistic Heavy Ion Collider (RHIC) and Large Hadron Collider (LHC) has been shown to behave like an almost perfect fluid, i.e. a fluid with little to no viscosity. Refs.\, \cite{Arsene:2004fa,Back:2004je,Adams:2005dq,Adcox:2004mh,ALICE:2011ab,ATLAS:2011ah,ATLAS:2012at,Chatrchyan:2012wg} are some of the first publications from each experimental collaboration at RHIC and LHC. Many more detailed studies of more complex observables have followed since and generally support the fluid interpretation strongly. See  \cite{Heinz:2013th,Gale:2013da,Gajdosova:2020nvb} for reviews.

This fluid is of the size of a nucleus, approximately $10^{-14}$ meters in diameter and reaches temperatures 100,000 times greater than those in the core of the sun. Measurements in smaller collision systems, namely proton+proton (p+p) \cite{Khachatryan:2010gv} and proton+lead (p+Pb) \cite{Abelev:2012ola,Aad:2012gla,CMS:2012qk}
collisions at LHC, as well as proton+gold (p+Au) \cite{Aidala:2016vgl,Adam:2019woz}, deuteron+gold (d+Au) \cite{Adare:2013piz,Adare:2014keg,Adamczyk:2014fcx,Adam:2019woz}, and $^3$He+Au \cite{Adare:2015ctn,Lacey:2020ime} collisions at RHIC, have shown similar behavior as that observed in heavy ion collisions, and have triggered a variety of new theory developments to understand if we are creating an even smaller fluid (approximately 10 times smaller in diameter) in these collisions, or whether other phenomena, such as color correlations of dense gluon fields in the incoming projectile and target, or quantum interference effects are the dominant source of the measured signals. See \cite{Dusling:2015gta, Loizides:2016tew, Schlichting:2016sqo, Nagle:2018nvi} for reviews. 
We will discuss these theory developments, separating them into three sections:

\emph{What is a fluid and how small can it be?  – The applicability of hydrodynamics} Hydrodynamic simulations with an appropriate initial state model for the fluctuating geometry do a surprisingly good job in describing the systematics of the measured azimuthal anisotropies in the transverse momentum distribution of produced particles in systems with only 10 or more charged hadrons produced per unit rapidity. These momentum anisotropies are the simplest and cleanest observables that have revealed fluid like behavior in heavy ion collisions. Consequently, a lot of recent effort has been invested into understanding how hydrodynamics can be a good description of rather dilute systems, and how quickly a system can approach \emph{hydrodynamization}, i.e., can reach a state in which hydrodynamics provides a proper description. This involves the study of hydrodynamic vs. non-hydrodynamic modes, hydrodynamic expansions in strong and weak coupling calculations, as well as explicit simulations of the Boltzmann equation or dynamics of shockwave collisions within the framework of Anti-de Sitter/Conformal Field theory correspondence (AdS/CFT). The upshot from these studies is that hydrodynamics provides a good description of a system for rather low numbers of particles and at times that are significantly earlier than isotropization or equilibration times. 

\emph{What is the shape of the smallest fluid?  – Subnucleonic structure} A variety of experimental data, ranging from diffractive vector meson production in electron+proton (e+p) collisions, to p+p collisions, to other small system and even heavy ion collisions, indicate that the proton’s shape itself fluctuates from event to event. We will review work that describes proton+nucleus (p+A) collisions using the Color Glass Condensate (CGC) Effective Field Theory (EFT) coupled to fluid dynamic simulations, which can only get close to the experimentally observed azimuthal momentum anisotropies of the produced charged hadrons when fluctuating subnucleonic structures of the proton are included. We will discuss how to constrain these fluctuations using incoherent diffractive vector meson production in e+p collisions, which was measured at the Hadron-Electron Ring Accelerator (HERA) at the Deutsches Elektronen-Synchrotron (DESY). Using these constrained fluctuating protons instead of previously assumed approximately spherical protons improves agreement with the experimental data significantly. We will further review how the subnucleon structure evolves with energy, which is calculable within the CGC EFT. This allows comparison to the center of mass energy dependence of cross sections measured at HERA, and is important to predict the collision energy and rapidity dependence of many observables in small system collisions at RHIC and LHC. We will further discuss other recent calculations that focus on either the role of color charge fluctuations or nucleon position fluctuations in heavy ion collisions.

\emph{More than a fluid – Initial state momentum anisotropies} While the case for the important role of final state effects in small system collisions that produce many particles is very strong, other sources of anisotropies have been increasingly discussed after measurements in small systems with very few particles have also revealed significant elliptic and higher order azimuthal anisotropies. These sources produce anisotropies that are generally already encoded in the initially produced particles, are uncorrelated (in some cases even weakly anti-correlated) with the initial geometry, and have both classical and quantum origins. We will summarize calculations that produce such anisotropies, with focus on the CGC EFT, which has been most widely used. We will discuss the inability of these calculations to describe all the systematics in the experimental data by themselves, and will close with recent developments that combine both initial and final state sources of anisotropies, and introduce measurements that can potentially separate the two in the experimental data.

Small system collisions at both RHIC and LHC have driven theoretical developments on a broad range of topics. Besides heavy ion physics, new insights into hydrodynamics as an effective theory will have relevance to any field where relativistic hydrodynamics is used, most notably in the description of high-energy astrophysical phenomena, such as supernovae, jets, and gamma-ray bursts \cite{Marti:1999wi, Font:2000pp}. The developments in understanding nuclear substructure and gluon dynamics in high energy hadrons and nuclei is and will be directly applicable to processes studied at HERA and to be explored in depth at a future Electron Ion Collider (EIC) \cite{Accardi:2012qut}.

\section{What is a fluid and how small can it be? -- The applicability of hydrodynamics}
In this review we shall call a fluid any system that is well described by (viscous) fluid dynamics (or more colloquially hydrodynamics). 
Hydrodynamics has been found to describe particle production in small system collisions well \cite{Bozek:2011if,Bozek:2012gr,Bozek:2013df,Bozek:2013uha,Bozek:2013uha,Bozek:2013ska,Bzdak:2013zma,Qin:2013bha,Werner:2013ipa,Kozlov:2014fqa,Romatschke:2015gxa,Shen:2016zpp,Weller:2017tsr}. This includes particle spectra as functions of transverse momenta, as well as more detailed observables, such as the so called flow harmonics $v_n$, which measure the azimuthal anisotropy of particle production in the plane transverse to the beam line.

However, one is naturally driven to ask whether it makes sense to believe that hydrodynamic behavior is achieved in a system that produces as little as 10 charged hadrons per unit rapidity. Considering that the applicability of hydrodynamics is debatable even in heavy ion collisions, the presence of a significant hydrodynamic phase in small system collisions may seem unlikely. In heavy ion collisions, the issue is our lack of a detailed understanding of how an initially extremely anisotropic and rapidly expanding system can approach local isotropy in momentum space \cite{Arnold:2004ti} (or even local thermal equilibrium) on a time scale that is at least an order of magnitude shorter ($\mathcal{O}(1\,{\rm fm}/c\approx 3.3\times 10^{-24}{\rm s})$) than the total lifetime of the system ($\mathcal{O}(10\,{\rm fm}/c)$). Calculations relying on microscopic theories in both the weak and strong coupling limits predict pressure anisotropies of order 50\%, meaning that the transverse pressure (perpendicular to the beam line) is on the order of two times larger than the longitudinal (along the beam direction) pressure, at a time of approximately $1\,{\rm fm}/c$ after the collision \cite{Romatschke:2016hle}.

Yet, hydrodynamics describes such anisotropic systems very well, despite the large anisotropy at early times \cite{Chesler:2009cy,Heller:2011ju,Casalderrey-Solana:2013aba,Kurkela:2015qoa,Keegan:2015avk}. In particular, as mentioned above, the pressure anisotropy of the system can be well approximated by hydrodynamics even when there is a factor of 2 difference between the longitudinal and transverse pressure in a rapidly longitudinally expanding system.
This means that before the system isotropizes or thermalizes (if it ever does) it \emph{hydrodynamizes}, i.e., can be described by hydrodynamics. In other words, there exists a non-equilibrium attractor for the energy momentum tensor (i.e., the energy momentum tensor evolves toward this attractor solution for a wide variety of initial conditions), which is well described by viscous hydrodynamics after a time of approximately $1\,{\rm fm}/c$ or less \cite{Heller:2013fn,Heller:2015dha,Buchel:2016cbj,Denicol:2016bjh,Heller:2016rtz,Florkowski:2017olj,Romatschke:2017vte,Bemfica:2017wps,Spalinski:2017mel,Romatschke:2017acs,Behtash:2017wqg,Florkowski:2017jnz,Florkowski:2017ovw,Strickland:2017kux,Almaalol:2018ynx,Denicol:2018pak,Behtash:2018moe,Strickland:2018ayk,Heller:2018qvh,Mazeliauskas:2018yef,Behtash:2019qtk,Strickland:2019hff,Jaiswal:2019cju,Kurkela:2019set,Chattopadhyay:2019jqj,Brewer:2019oha,Kurkela:2019kip}. For a detailed review see \cite{Florkowski:2017olj}.

A way to understand the phenomenon of hydrodynamization, i.e., the applicability of hydrodynamics in systems far from equilibrium, is to consider the various modes, which can be either transient modes or long-lived hydrodynamic modes. In a typical situation, as time evolves transient modes decay exponentially fast (on a time scale that depends on the details of the microscopic theory), hydrodynamic modes begin to dominate, and one approaches a quasi-universal attractor behavior (We note that there are also far-from-equilibrium, early-time attractors, that can be reached prior to the late time hydrodynamic ones \cite{Kurkela:2019set,Almaalol:2020rnu}). So it is likely the case (the universality of this statement has not yet been proven) that approaching the attractor is equivalent to achieving ``hydrodynamic behavior" \cite{Florkowski:2017olj}. This means that determination of the applicability of hydrodynamics should not rely on the smallness of subsequent terms in a gradient expansion, because it is divergent (because of the presence of non-hydrodynamic modes) \cite{Heller:2013fn,Heller:2015dha}, but instead the dominance of hydrodynamic modes.

Based on the momentum scales at which hydrodynamic modes vanish completely from the spectrum, one can even attempt to estimate the minimal size of a droplet, for which hydrodynamics can apply (i.e., one estimate for "the smallest fluid on earth"). It was argued to be $\sim 0.15\,{\rm fm}=0.15\times 10^{-15} {\rm m}$ in \cite{Romatschke:2016hle}. In terms of the global size of systems produced in realistic small system collisions (typically of the proton size scale and larger), this is a small scale, however, one should keep in mind that when including subnucleonic structure (see Sec.\,\ref{sec:subnucleon}), local hot spots can be on the order of this size scale.

One should note that different viscous hydrodynamic schemes, such as Mueller Israel Stewart (M-IS) \cite{Muller:1967zza,Israel:1976tn, Israel:1979wp,Muller:1999in}, Baier Romatschke Son Starinets Stephanov (BRSSS) \cite{Baier:2007ix}, Denicol Niemi Molnar Rischke (DNMR) \cite{Denicol:2012cn}, or anisotropic hydrodynamics \cite{Florkowski:2010cf,Martinez:2010sc,Ryblewski:2010bs,Florkowski:2013lza,Nopoush:2014pfa,Nopoush:2014qba,McNelis:2018jho} (see \cite{Strickland:2014pga,Alqahtani:2017mhy} for reviews and more references), all have different transient modes. How well they describe a given microscopic theory on short time scales depends on how well they reproduce the transient modes of that theory. 
In many cases, anisotropic hydrodynamics (both the leading order kind, which assumes a spheroidal particle distribution function, and even more so the next to leading order implementation, which allows additional arbitrary viscous corrections to this distribution function) has succeeded best in reproducing results from microscopic kinetic theories, in particular the time evolution of the diagonal components of the stress energy tensor when the shear viscosity to entropy density ratio is large (compared to the strong coupling limit of $\eta/s=1/4\pi$)  \cite{Florkowski:2013lza,Nopoush:2014pfa,Nopoush:2014qba,Martinez:2017ibh,Chattopadhyay:2018apf}. 

The differences between different hydrodynamic schemes could well be important in small collision systems, where the total lifetime of the fireball is often less than $3\,{\rm fm}/c$. However, before one can make any conclusive statements about which hydrodynamic scheme is superior for describing nuclear collisions, one needs to better understand the underlying microscopic non-equilibrium theory in the first place.  Recent progress on that front includes the study of non-equilibrium dynamics in weak coupling asymptotics \cite{Berges:2013fga}, which shows consistency with the bottom-up thermalization scenario \cite{Baier:2000sb}, and the implementation of effective QCD kinetic theory \cite{Arnold:2002zm} to describe the early time evolution before hydrodynamization \cite{Kurkela:2018vqr,Kurkela:2018wud}. The latter is a good description for a more dilute situation, where quantum effects are important, and which is reached rather rapidly via expansion, even when starting with an overoccupied system described by classical fields and the Yang-Mills equations (like in the IP-Glasma model \cite{Schenke:2012wb,Schenke:2012hg}). Going from one limit to the other smoothly is possible because there is a regime of occupation where both descriptions (classical Yang-Mills fields and distribution functions of kinetic theory) are valid \cite{Kurkela:2016vts}. 

On a microscopic level, the coupling in the kinetic theory (or alternative) description must be strong enough to overcome the rapid expansion of the system and approach the hydrodynamic attractor for a system with a realistic small shear viscosity to entropy density ratio of several times $1/4\pi$. Therefore, calculations in kinetic theory are performed at weak coupling and then usually extrapolated to values of the coupling that are compatible with the proper hydrodynamic description \cite{Kurkela:2018vqr,Kurkela:2018wud,Kurkela:2020wwb}. There are also indications \cite{Giacalone:2019ldn,Kamata:2020mka} that different underlying theories produce similar background evolution and response functions as used in the effective kinetic theory description of \cite{Kurkela:2018vqr,Kurkela:2018wud}, such that the approach to the attractor may be well approximated even if the true underlying theory is not perfectly captured by the effective kinetic model.

Consequently, one strategy for a complete description of a heavy ion (or small system) collision is to start with a classical framework, such as the IP-Glasma model, couple it to an effective kinetic theory description, and then transition to viscous hydrodynamics (followed eventually by microscopic hadronic transport as the system becomes dilute again at late times). Including the proper early time non-equilibrium transport phase could have a strong effect on the production of electromagnetic probes in small systems, as they are produced throughout the entire evolution, and once produced are no longer modified. Investigations of non-equilibrium photon production is ongoing \cite{Kasmaei:2019ofu,Gale:2020dum,Churchill:2020uvk} and could shed more light on the details of the complex early time dynamics, especially in small systems. Furthermore, photons and dileptons should also be sensitive to chemical equilibration (between quark and gluon degrees of freedom) \cite{Kurkela:2018xxd,Kurkela:2018oqw,Du:2020dvp,Du:2020zqg}, as delayed equilibration would suppress the early time photon and dilepton yield (gluons will dominate for longer), affecting both spectra and $v_n$ of these electromagnetic probes \cite{Gale:2020xlg}.

In conclusion, recent theoretical developments have demonstrated that hydrodynamics can describe systems that are out of equilibrium, even when the pressure anisotropy is on the order of a factor two. The criterion for the applicability of a hydrodynamic expansion should not be the smallness of subsequent terms in the expansion, as the series is not convergent anyways, but the dominance of hydrodynamic modes over transient modes. This dominance typically sets in well before isotropization is achieved, and is dubbed "hydrodynamization". This behavior can explain the naively surprising success of hydrodynamic models in describing experimentally observed bulk properties of small system collisions, such as d+Au collisions at RHIC, p+Pb collisions at the LHC, and many others. Nevertheless, significant uncertainty remains in our understanding of the earliest times during the system's evolution - so far only approximate solutions for the correct microscopic description in either the weak or strong coupling limits have been explored. Also, effective kinetic descriptions that describe the early time evolution and eventually connect to hydrodynamics have so far only been considered in the conformal limit. It will be important, both from a theory and a phenomenological perspective, to extend these studies to non-conformal systems, as early time bulk viscous effects could play an important role for a variety of observables \cite{NunesdaSilva:2020bfs}. Finally, it remains to be investigated how the presence of spatial fluctuations (see Sec.\,\ref{sec:subnucleon}) affects the conclusions on hydrodynamization in both kinetic theory and AdS/CFT frameworks.

Despite these remaining challenges, the progress made is significant. The theory of viscous relativistic hydrodynamics has evolved tremendously over the last several years, and it is driven predominantly by the field of heavy ion and small system collisions. These developments will likely have impacts far beyond this field, including for example the description of neutron star mergers \cite{Duez:2004nf,Shibata:2017xht,Most:2018eaw,Alford:2019kdw} and condensed matter systems \cite{Sachdev:2010ch,Nastase:2017cxp}. For detailed recent reviews on the question of thermalization in QCD as well as hydrodynamization, we refer the interested reader to \cite{Schlichting:2019abc,Berges:2020fwq}.

\section{What is the shape of the smallest fluid? -- Subnucleonic structure}\label{sec:subnucleon}
The anisotropy in the particle production transverse to the beam line is characterized by Fourier coefficients $v_n$, which represent the amplitudes of the $\cos(n\phi)$ modulation, with $\phi$ being the azimuthal angle and $n$ an integer. In the hydrodynamic framework, these anisotropies are generated by the response of the strongly interacting system to the initial geometry in the transverse plane of the collision. 
In heavy ion collisions, odd flow harmonics, such as $v_3$, $v_5$, etc., \footnote{Directed flow $v_1$ has a rapidity even component, that is entirely driven by fluctuations, but it also has a rapidity odd component, that is finite even in the case of no fluctuations.} along with the details of the even harmonics, are driven by fluctuations in the initial geometry \cite{Alver:2010gr} (the average geometry has a symmetry that allows only even harmonics). The details of these fluctuations modify the event by event distribution of the flow harmonics, and by that their cumulants, which are accessible by studying multiparticle correlations, e.g. of the produced charged hadrons \cite{Borghini:2000sa, Borghini:2001vi, Bilandzic:2013kga}.
Similarly, if dominated by the same process, namely the final state response to the initial geometry, $v_n$ should also be driven by fluctuations in small systems. In this case, considering for example p+A collisions, also the even harmonics are strongly dominated by event by event fluctuations, as the average ellipticity is close to zero for central collisions.

\begin{figure}[tb]
\centering
\includegraphics[width=0.49\linewidth]{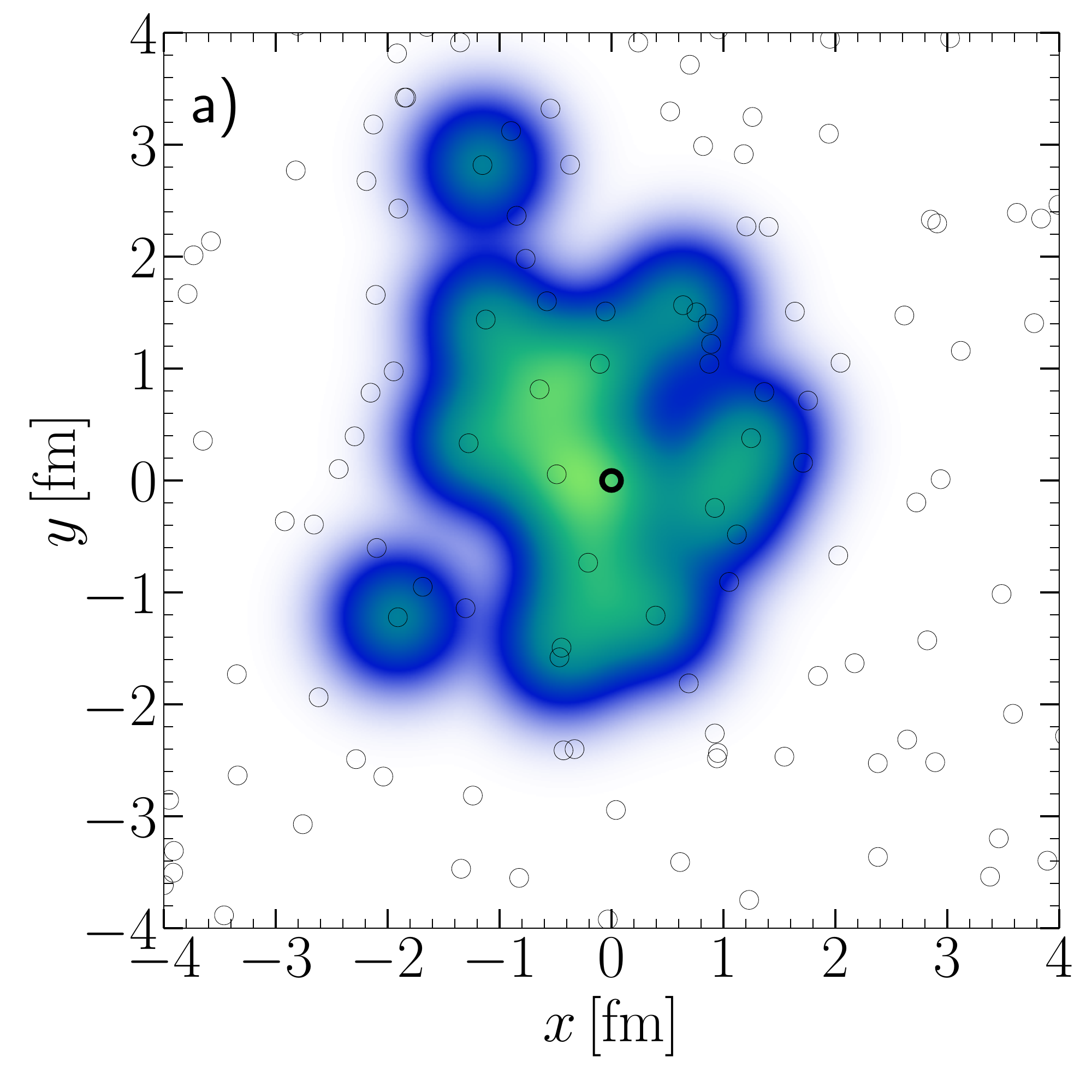}
\includegraphics[width=0.49\linewidth]{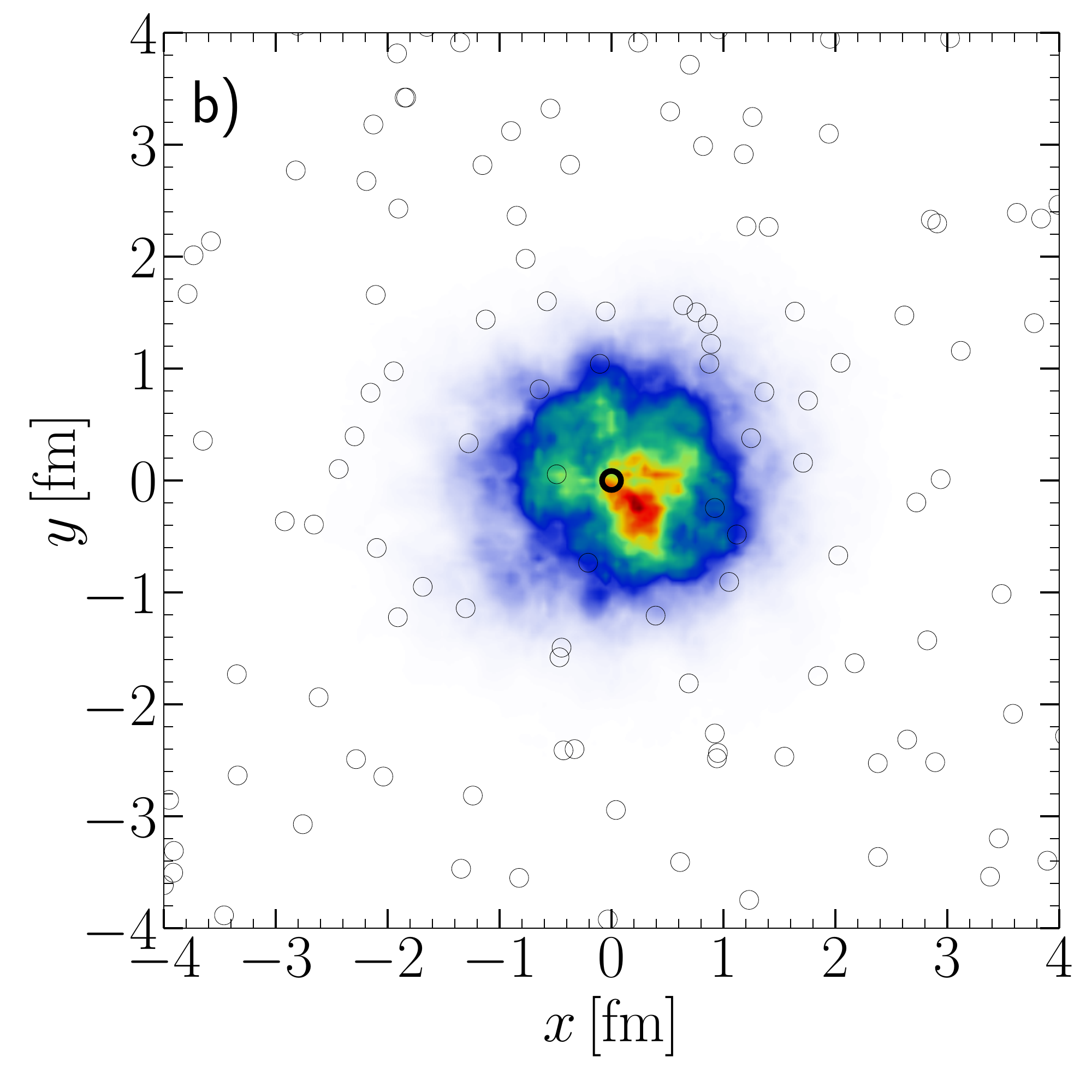}
\caption{Profile of the energy density distribution (normalization arbitrary) in the transverse plane in a single $5.02\,{\rm TeV}$ p+Pb collision computed in a) a Monte-Carlo Glauber model using nucleon degrees of freedom, where the energy density is proportional to the sum of thickness functions of all wounded (participating) nucleons, and b) the IP-Glasma model assuming round nucleons. Thin circles represent the nucleon positions in the Pb nucleus, the thick circle is the proton projectile position. Nucleon positions are the same for both cases a) and b). \label{fig:models}}
\end{figure}

When using fluctuating Monte Carlo Glauber type models \cite{Miller:2007ri,dEnterria:2020dwq} to initialize small system collisions, nucleon degrees of freedom (i.e., using nucleons without any substructure) have been shown to provide sufficient fluctuations to produce azimuthal anisotropies close to those measured in experiments at RHIC and LHC \cite{Bozek:2011if,Bozek:2012gr,Bozek:2013df,Bozek:2013uha,Bozek:2013uha,Bozek:2013ska,Bzdak:2013zma,Qin:2013bha,Werner:2013ipa,Kozlov:2014fqa,Romatschke:2015gxa,Shen:2016zpp,Weller:2017tsr,Giacalone:2017uqx}. This is possible only for certain choices for the energy or entropy deposition. With a proton projectile, assuming a spherical proton, fluctuations can only originate from the fluctuating positions of nucleons in the heavy ion target. Thus, energy (or entropy) deposition that is proportional to a sum of the projectile $(A)$ and target's $(B)$ wounded nucleon nuclear matter densities, characterized by the thickness functions ($\sim T_A+T_B$)\footnote{The thickness function $T_A$ is defined as the integral over the three dimensional nuclear density distribution along the direction of the beam line.} is able to produce rather lumpy structures in the transverse plane of the collision (with fluctuations on the nucleon size scale), which is necessary to produce the large $v_n$ observed experimentally (See Fig.\,\ref{fig:models} a)).

\begin{figure}[tb]
\centering
\includegraphics[width=0.6\linewidth]{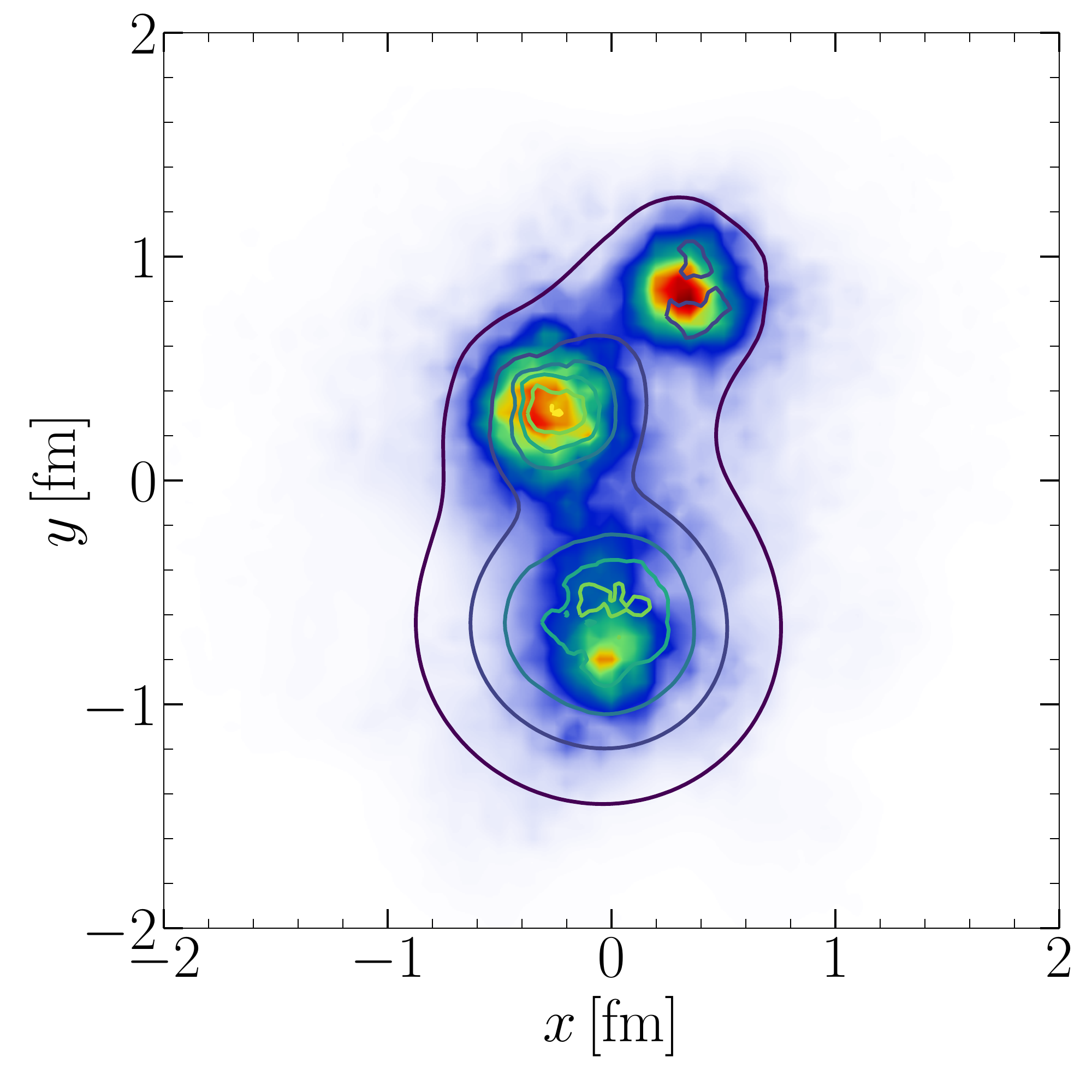}
\caption{The color map shows the energy density distribution (arbitrary units) in the plane transverse to the beam direction for a p+Pb collision. The contour lines indicate the shape of the projectile proton (quantified using a measure of the gluon density in the proton). \label{fig:proton_eden}}
\end{figure}

However, several arguments disfavor this type of energy deposition. First, in a model that parametrizes the functional dependence of the deposited entropy distribution on the thickness functions (\trento{}) \cite{Moreland:2014oya}, the consistent outcome of a range of Bayesian analyses has been that the experimental data prefers an entropy density proportional to the square root of the product of the two thickness functions \cite{Bernhard:2016tnd,Everett:2020yty,Everett:2020xug,Nijs:2020roc}. In small systems, in particular p+A collisions, this would lead to too little fluctuations in the transverse geometry, such that subnucleon structure was argued to be essential in order to describe both p+Pb and Pb+Pb systems simultaneously using this initial state model \cite{Moreland:2018gsh}. 

Furthermore, the IP-Glasma initial state model \cite{Schenke:2012wb,Schenke:2012hg}, that is based on an actual effective theory of QCD, the CGC \cite{McLerran:1994ni,McLerran:1994ka,Kovner:1995ja,Iancu:2003xm}, predicts that the initial energy density is proportional to the product of the two thickness functions. While slightly different from the \trento{} model result, such a dependence will also lead to extremely small spatial fluctuations in the initial geometry when one assumes a round proton, see Fig.\,\ref{fig:models} b). This, along with the resulting small azimuthal anisotropies, was shown explicitly in \cite{Schenke:2014zha}. 
Interestingly, a similar dependence on the thickness functions as in this weak coupling limit calculation is also obtained in the extremely strong coupling limit \cite{Romatschke:2013re,Romatschke:2017ejr} using AdS/CFT correspondence \cite{Maldacena:1997re}.

If the energy deposition is proportional to the product of thickness functions, geometric fluctuations large enough to generate as much anisotropic flow as observed in the experimental data on p+A collisions can only be generated when including a nucleonic substructure. Otherwise, the overlap region would take on the shape of the mostly round projectile proton (in IP-Glasma there are small color charge fluctuations, which produce some small geometric anisotropies, but they would not be sufficient to generate large anisotropic flows). We illustrate how the energy density distribution in the transverse plane for a p+Pb collision follows the shape of the proton projectile in Fig.\,\ref{fig:proton_eden}. The contour lines indicate a measure of the gluon density in the incoming proton. The energy density, represented by the color profile resembles this shape closely, with variations resulting from fluctuations in the lead target.

Other indications of the relevance of a subnucleonic structure include studies by the PHENIX Collaboration \cite{Adler:2013aqf}, where it was found that particle production in asymmetric (small) and symmetric collisions can all be explained by assuming constituent quark degrees of freedom \cite{Eremin:2003qn} (also see \cite{Bialas:1980zw}), while using nucleon participants $N_{\rm part}$ and number of nucleon-nucleon collisions $N_{\rm coll}$ in the well known form $[(1-x)\langle N_{\rm part}\rangle/2+x \langle N_{\rm coll}\rangle]$ (with a free parameter $x$) only works well to describe particle production in symmetric systems \cite{Adler:2013aqf}. 
Additionally, because particle production in the constituent quark picture does not explicitly depend on $N_{\rm coll}$, it leads to a better description of the correlation of $v_2$ and $N_{\rm ch}$ in ultracentral U+U and Au+Au collisions, compared to that obtained with the two-component Glauber ansatz \cite{Adamczyk:2015obl}.

For p+p collisions, it was shown \cite{Albacete:2016pmp} that the hollowness effect \cite{Arriola:2016bxa,Alkin:2014rfa,Dremin:2015ujt,Troshin:2016frs}, which refers to the inelasticity density of the collision not reaching its maximum at zero impact parameter and was observed in p+p collisions at $\sqrt{s} = 7\,{\rm TeV}$ \cite{Antchev:2011zz}, can also be explained when considering subnucleonic hot spots in the proton. Finally, there are indications for size fluctuations of the proton in jet measurements in p+Pb or d+Au collisions as a function of centrality \cite{McGlinchey:2016ssj}.

So there are plenty of indications for the relevance of a fluctuating nucleon substructure, but the details of what this structure looks like are less clear. 
As there are no first principles calculations available yet, that would determine the fluctuating geometric structure of a nucleon, there are two possible ways to proceed. First, one can parametrize the substructure, typically as a combination of $N$ hot spots with variable distributions of their width and position within the nucleon. Then, to constrain the parameters one can perform a Bayesian analysis of heavy ion and small system collisions and this way optimize the agreement with experimental data. This procedure was adopted in \cite{Moreland:2018gsh} (also see \cite{Nijs:2020roc}). In this particular work, the number of subnucleonic hot spots could not be very well constrained, but a number greater than one (one corresponds to no substructure) was favored. The widths of the subnucleonic hot spots was on the other hand tightly constrained to approximately $0.4\,\rm{fm}$, close to what is typically used for the \emph{nucleon} width in the IP-Glasma model. 

Alternatively, one can attempt to constrain the subnucleon structure using independent measurements, e.g. from e+p collisions. As it was demonstrated in \cite{Mantysaari:2016ykx}, there is an exclusive process in e+p collisions that is particularly sensitive to geometrical fluctuations of the gluon distribution in the proton, namely the incoherent diffractive production of vector mesons: By Fourier conjugation of the transverse momentum variable, diffractive vector meson production provides information on the spatial gluon distribution in the proton target. In the case of coherent production, i.e., the case where the proton stays intact, the $|t|$ (where $|t|$ is the transverse momentum transfer squared) contains information on the average size of the target, but more interesting for our purpose, the incoherent differential cross section is proportional to the variance of the scattering amplitude, making it sensitive to its fluctuations, including those of geometric nature. 

\begin{figure}[tb]
\centering
\includegraphics[width=0.49\linewidth]{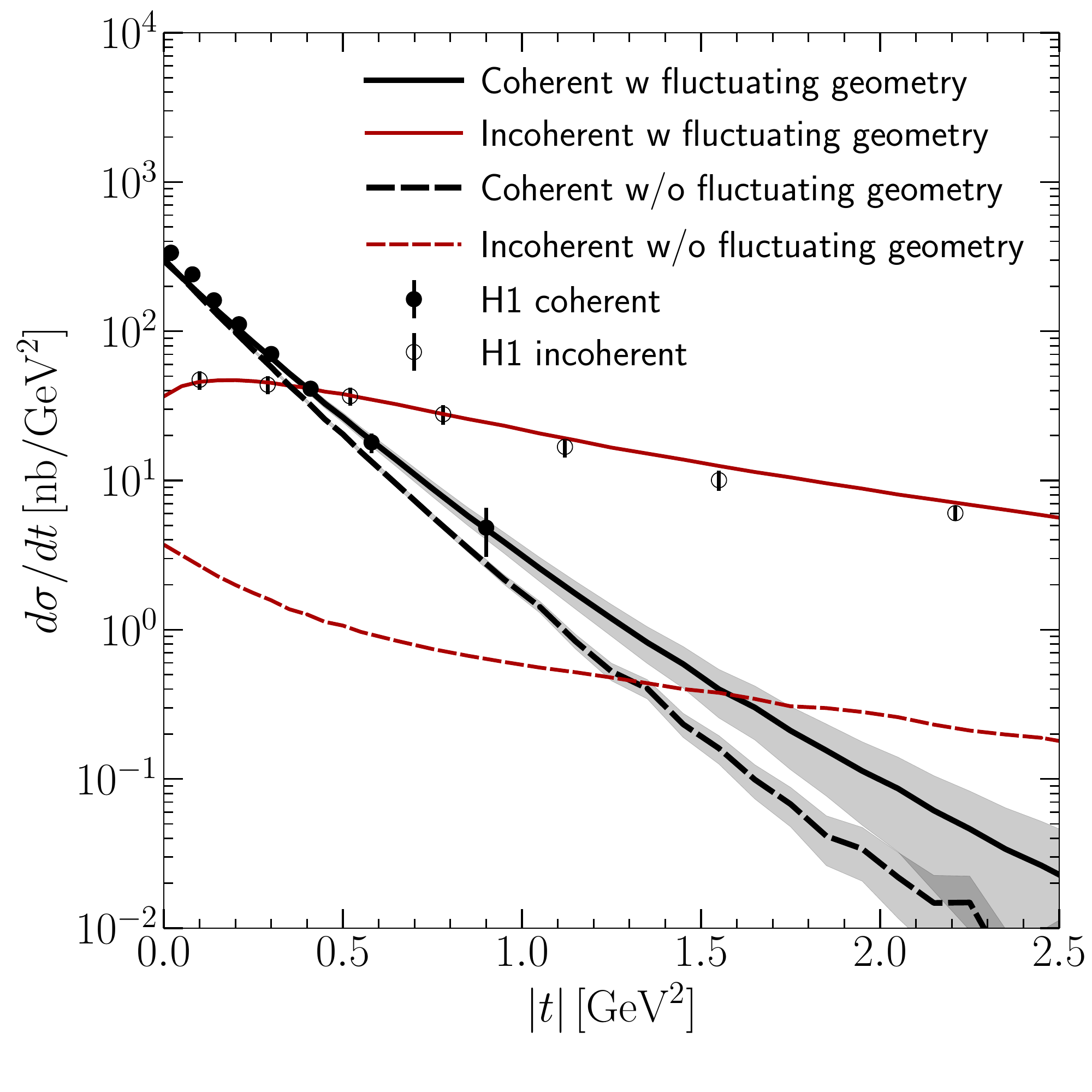}
\includegraphics[width=0.5\linewidth]{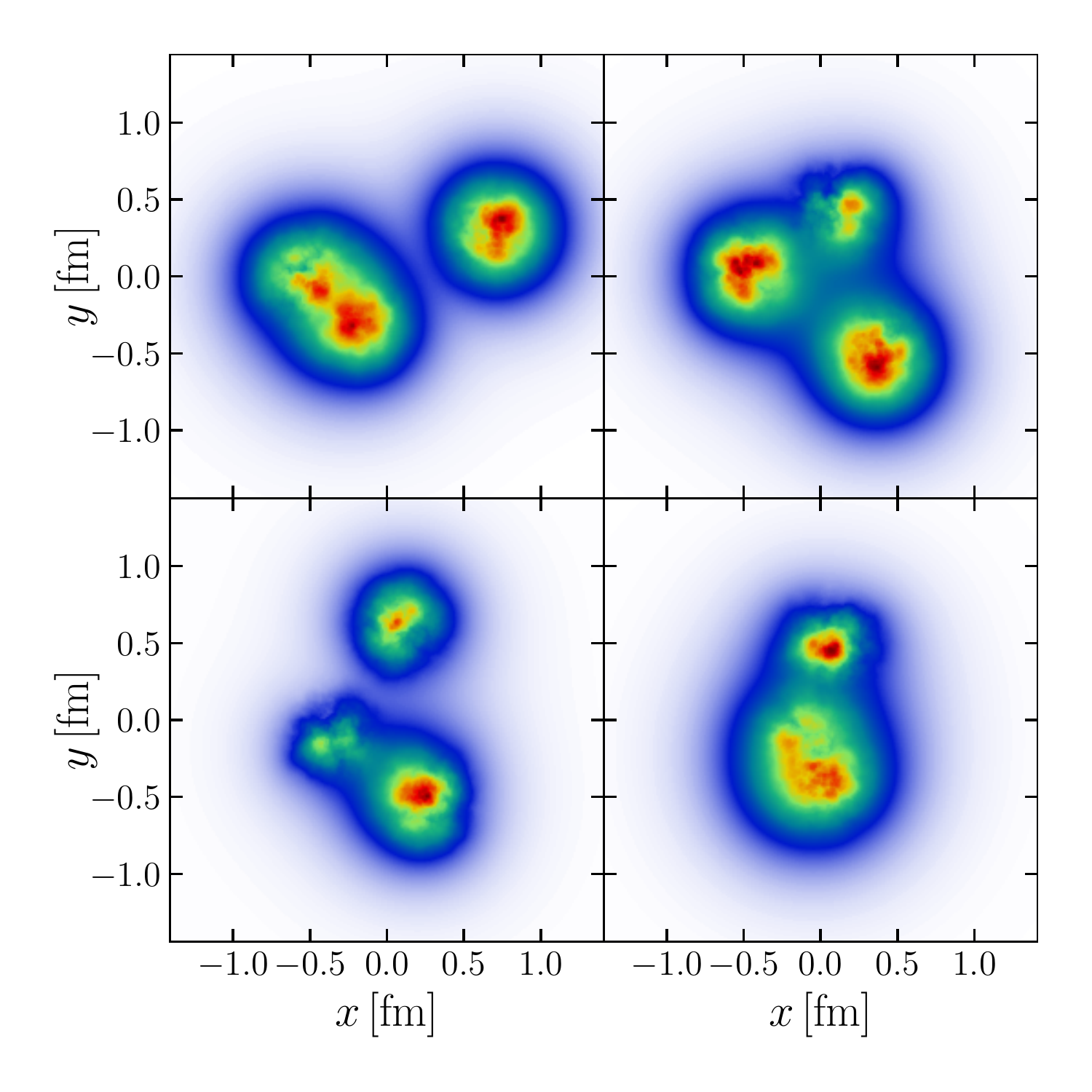}
\caption{Left: Coherent (black) and incoherent (red) production of $J/\psi$ vector mesons in e+p collisions 
\cite{Mantysaari:2016ykx,Mantysaari:2016jaz} compared to measurements from the H1 Collaboration \cite{Alexa:2013xxa}. Solid lines represent calculations including a fluctuating geometric substructure of the proton with three hot spots, dashed lines assume a round geometry and only include color charge fluctuations. Right: Four examples of proton configurations illustrating the degree of fluctuations necessary to describe the incoherent data on the left. Shown is (one minus) the real part of the trace of the proton's Wilson lines (normalized by $1/N_c$), which can be interpreted loosely as a gluon density.  \label{fig:JPsi}}
\end{figure}

 Motivated by discussions in \cite{Miettinen:1978jb,Frankfurt:1993qi,Frankfurt:2008vi,Caldwell:2009ke,Lappi:2010dd}, it was thus suggested in \cite{Mantysaari:2016ykx} to use data on diffractive $J/\psi$ production in e+p collisions at HERA \cite{Chekanov:2002rm,Aktas:2003zi,Aktas:2005xu,Chekanov:2002xi,Alexa:2013xxa} to constrain the proton average shape and fluctuating substructure within the IP-Glasma model (see also \cite{Mantysaari:2020axf}). The sensitivity of the incoherent differential cross section to the substructure turned out to be rather dramatic. As shown in Fig.\,\ref{fig:JPsi} (left), while the coherent diffractive cross section can always be approximately described by adjusting the average shape, describing the experimentally determined incoherent diffractive cross section requires the presence of substantial geometric fluctuations. Comparing to the result that assumes a Gaussian thickness function of the proton and includes only color charge fluctuations, the result obtained using three Gaussian hot spots (whose radius is approximately three times smaller than the proton radius) produces an incoherent cross section that is significantly larger, and also has a shape in $|t|$ that is much closer to that of the experimental data.\footnote{Many different models for subnucleonic spatial fluctuations in the proton are conceivable. For example, one could base a model on the spin fluctuations in the proton, as done in \cite{Miller:2003sa,Habich:2015rtj}.}  Three hot spots are motivated by the presence of three valence quarks, around which one assumes the gluons to be clustered. Other numbers of hot spots are certainly conceivable, as there could be more large $x$ degrees of freedom, such as large $x$ gluons or sea quarks, around which smaller $x$ gluons can cluster. An additional fluctuation of the normalization for each hot spot is also included, which mainly affects the low $|t|$ part of the incoherent spectrum.
 
As an illustration of the degree of fluctuations required by the incoherent diffractive HERA data, we also show four example protons in Fig.\,\ref{fig:JPsi}, visualized by plotting the real part of the trace of the gluon Wilson lines in the transverse plane (divided by $N_c$), which can be loosely taken to represent the density of gluons. The pictures reveal the three hot spots that have both fluctuating intensity and positions. We stress that the length scale for the subnucleon hot spots is not derived from theory, but extracted from experimental data. It is not clear exactly what sets this intermediate scale (between $1/\Lambda_{\rm QCD}$ and $1/Q_s$, where $Q_s$ is the saturation scale), but it must emerge from the dynamics of quarks and gluons at intermediate $x$. Perhaps lattice QCD calculations will be able to address this question in the future.

\begin{figure}[t]
\centering
\includegraphics[width=0.39\linewidth]{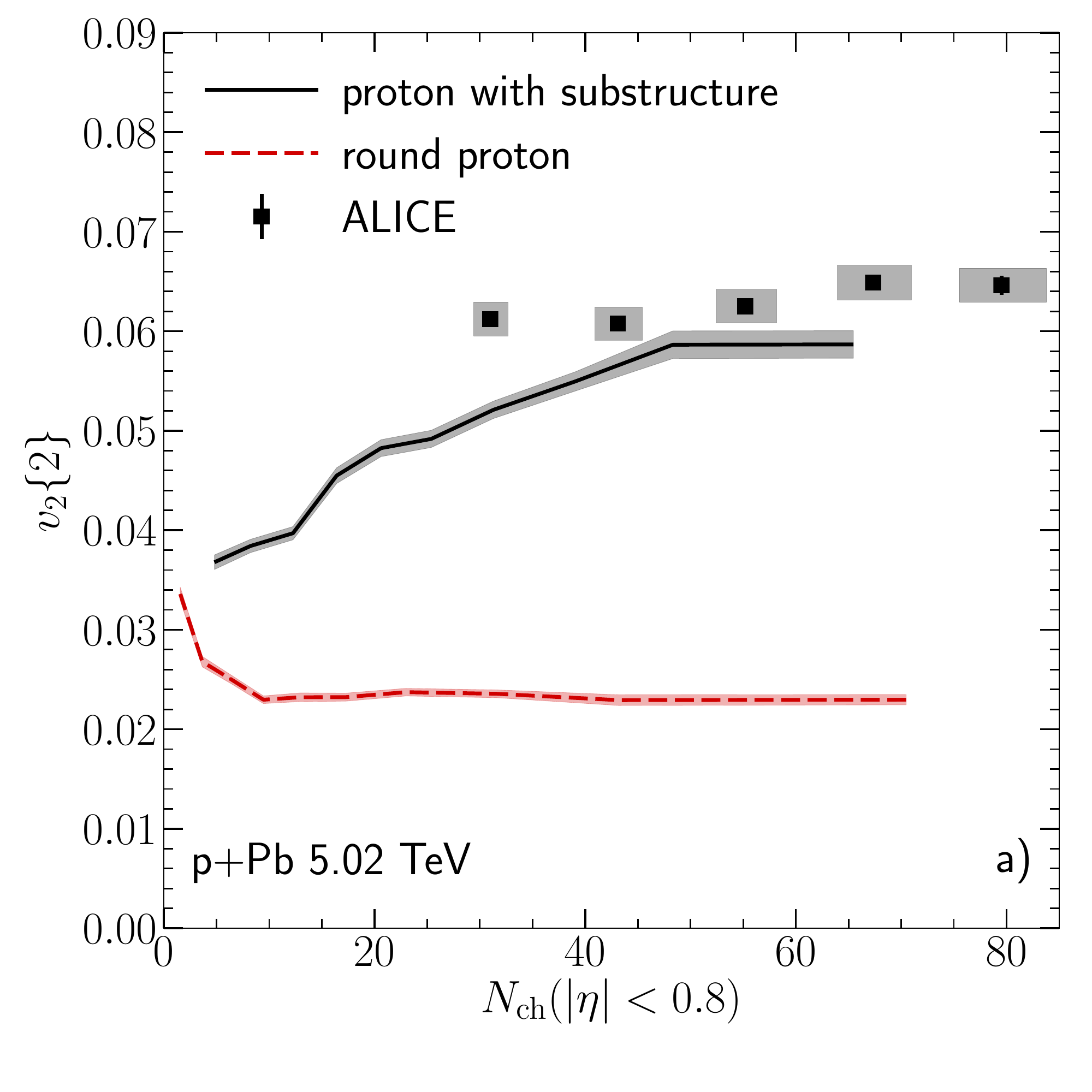}
\includegraphics[width=0.39\linewidth]{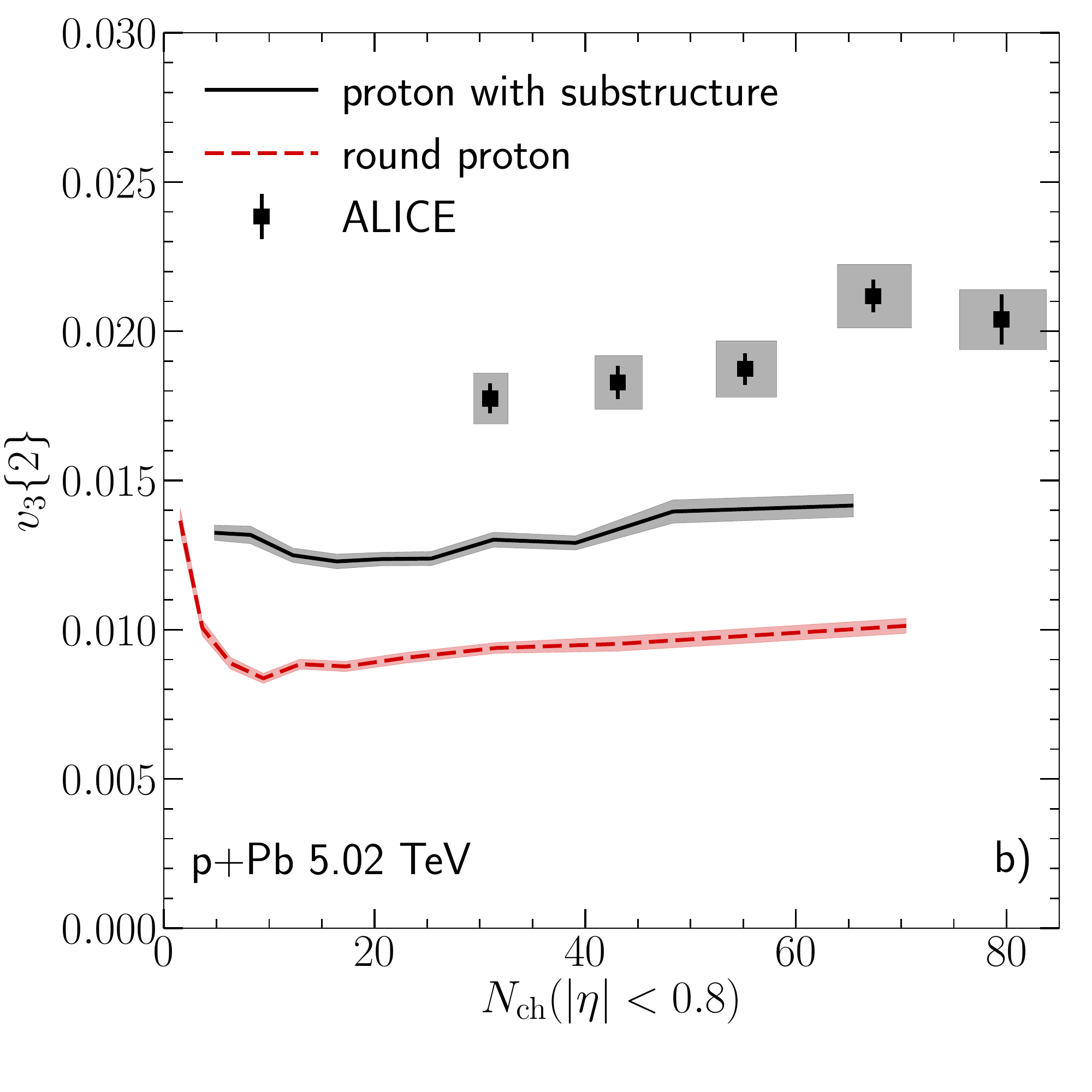}
\includegraphics[width=0.2\linewidth]{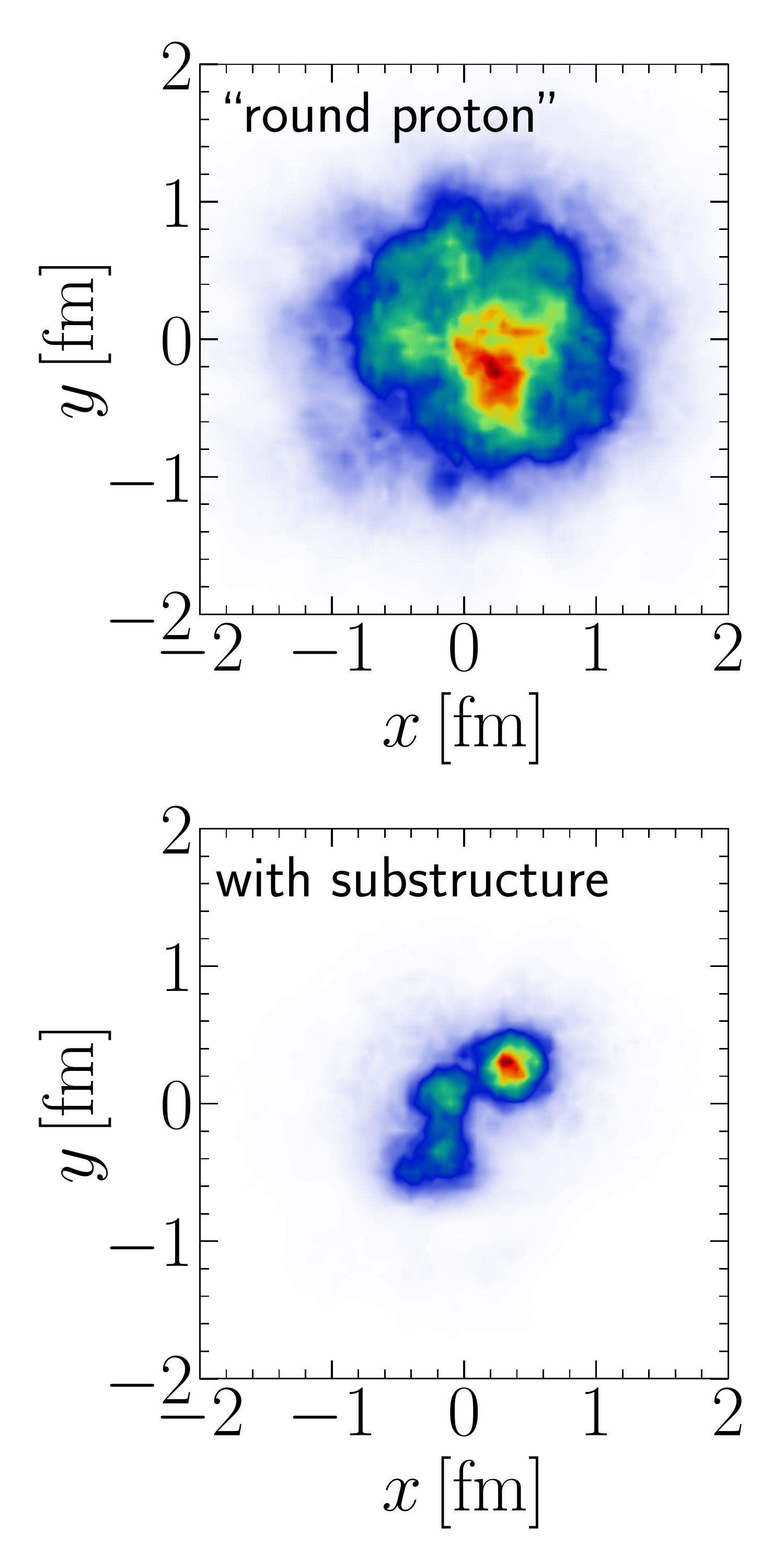}
\caption{The charged hadron azimuthal anisotropy harmonics from two particle correlations $v_2\{2\}$ (a) and $v_3\{2\}$ (b) computed within the IP-Glasma+\textsc{Music}+UrQMD hybrid model assuming a round thickness function for the proton (dashed lines) or proton substructure with three hot spots in the gluon distribution (solid lines). The increase of the $v_n$, driven by the increased fluctuations and larger eccentricities, is particularly dramatic for $n=2$.  Experimental data from the ALICE Collaboration \cite{Acharya:2019vdf}. On the right, we show the energy density distribution in the transverse plane (arbitrary units) for a typical collision with ``round'' nucleons on the top and one example collision using nucleons with substructure on the bottom. \label{fig:v2round}}
\end{figure}

One can now explore the effect of including these constrained subnucleon fluctuations in the initial condition of a hydrodynamic simulation of p+Pb collisions for example. Since above calculations of diffractive $J/\psi$ production were done in a framework that is identical to the pre-collision stage of the IP-Glasma model, the implementation in IP-Glasma is trivial. A full calculation using the constrained subnucleonic fluctuations was first done in \cite{Mantysaari:2017cni} where significantly larger $v_n$ compared to those for round protons \cite{Schenke:2014zha} were found. To demonstrate the increase of the $v_n$ and the improvement when comparing to experimental data, we show the results for the charged hadron $v_2\{2\}$ and $v_3\{2\}$ (where the $\{2\}$ indicates a root-mean-square measure obtained from 2-particle correlations) in 5.02 TeV p+Pb collisions from \cite{Schenke:2020mbo}, which uses a fluctuating nucleon structure together with the same calculation but assuming round nucleons in Fig.\,\ref{fig:v2round}.

The difference between the two cases is dramatic, especially for $v_2$, where for the most central events shown it is approximately a factor of 2.5. For $v_3$ the increase from including subnucleon structure is approximately a factor of 1.5. In the case of $v_3$ the experimental data is still underpredicted. One should note that the p+Pb result is a true prediction, as tuning of the model was only done for Au+Au collisions at RHIC energies. It is thus conceivable that a better description could be achieved when including all systems in a coordinated tune of the model. Nevertheless, the importance of subnucleon fluctuations is clearly demonstrated here, and, at least when focusing on $v_2$, p+Pb collisions at the LHC and diffractive vector meson production in e+p collisions seem to favor a similar degree of subnucleonic fluctuations in the proton.

As an illustration, Fig.\,\ref{fig:v2round} also shows the energy density distribution in the transverse plane for a typical central p+Pb collision with ``round'' nucleons on the top right and one example collision using nucleons with substructure on the bottom right. Eccentricities in the latter case are significantly larger compared to the round nucleon calculation, explaining the increase in the $v_n$ when including subnucleonic fluctuations.

\begin{figure}[tb]
\centering
\includegraphics[width=0.49\linewidth]{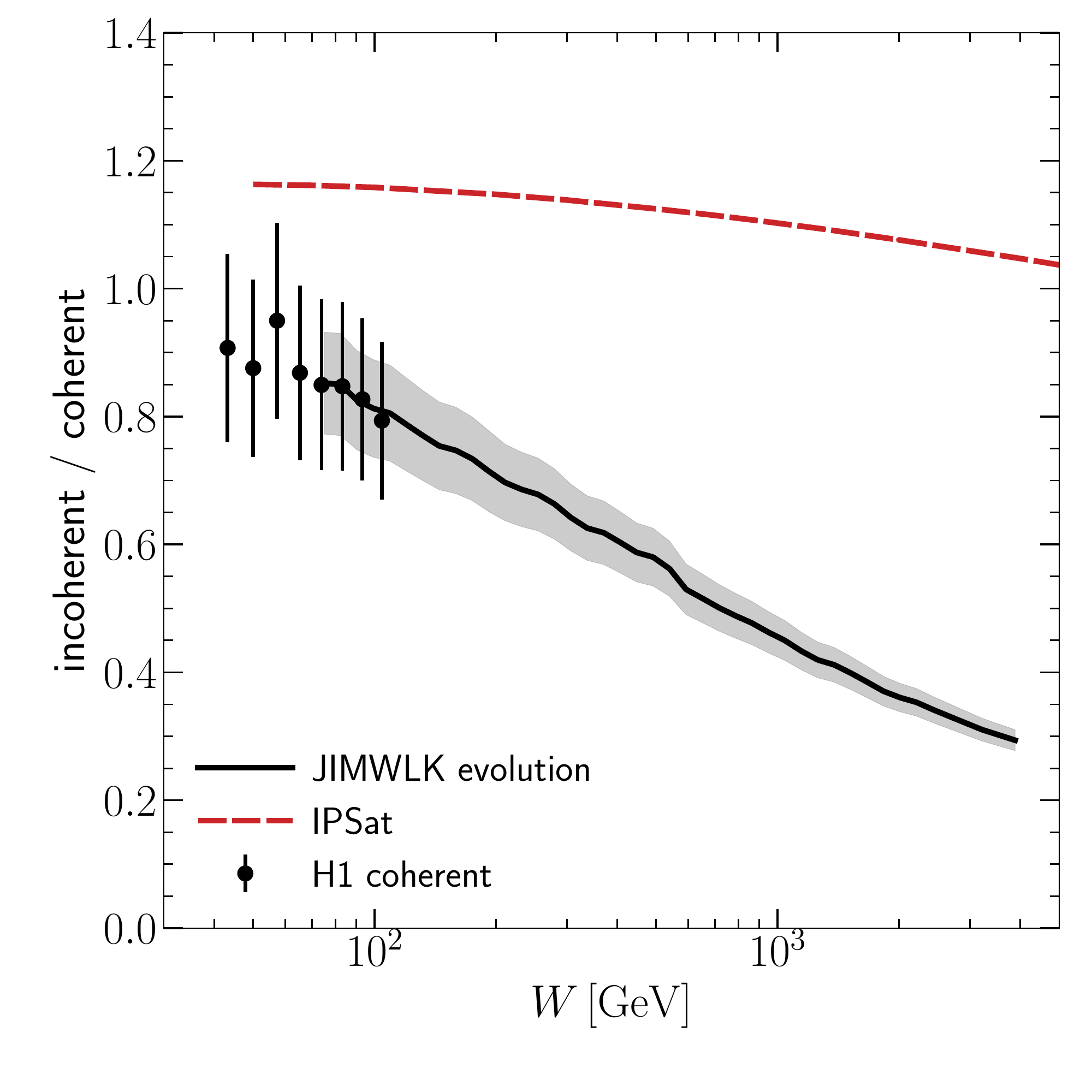}
\includegraphics[width=0.5\linewidth]{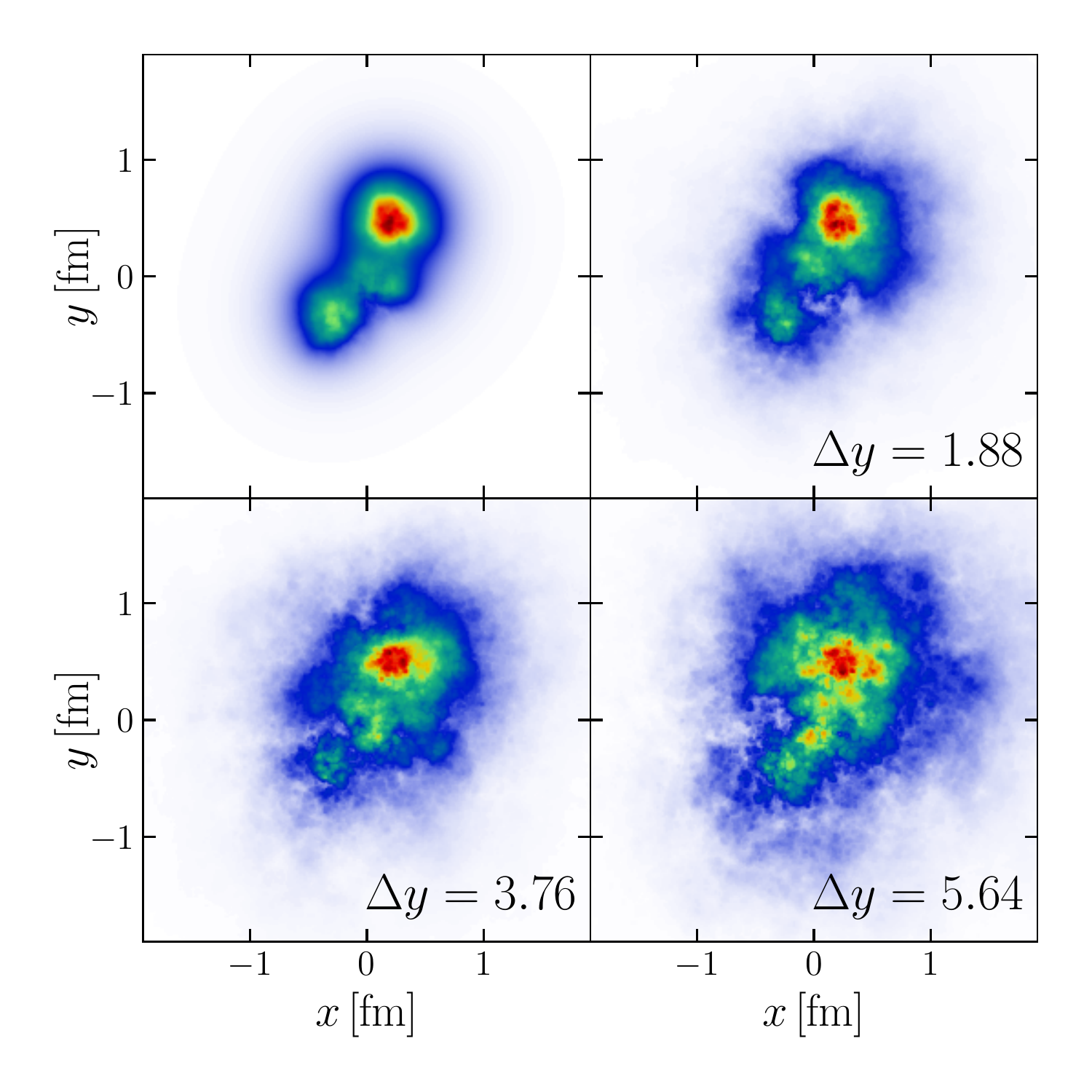}
\caption{Left: The ratio of the $W$ dependent incoherent to coherent cross sections from \cite{Mantysaari:2018zdd} compared to H1 data \cite{Alexa:2013xxa}. The solid line is a CGC calculation for fluctuating protons including JIMWLK evolution, the dashed line is the IPSat model calculation, which does not include the evolution of the proton size or the fluctuation scale. Right: The energy (rapidity, or $x$) evolution of one example configuration of a proton. $\Delta y$ indicates the evolution in rapidity from the initial configuration shown in the upper left. Shown is (one minus) the real part of the trace of the proton’s Wilson lines (normalized by $1/N_c$).\label{fig:jimwlk}}
\end{figure}

With the Electron Ion Collider (EIC) \cite{Accardi:2012qut,Aschenauer:2017jsk} on the horizon, 
more precise measurements of diffractive vector meson production have the potential to better constrain the subnucleonic structure not only of protons, but also light \cite{Mantysaari:2019jhh} and heavy nuclei (c.f. \cite{Mantysaari:2017dwh}, where ultraperipheral Pb+Pb collisions at LHC are discussed and can provide similar information). For e+p collisions it was recently suggested \cite{Mantysaari:2020lhf} to also measure the dependence of the $|t|$ differential \emph{incoherent} diffractive cross section on the azimuthal angle between the produced vector meson and the scattered electron, to get an additional handle on the substructure fluctuations. Of particular interest would also be the energy ($W$, the center-of-mass energy in the
virtual photon-proton scattering process), or Bjorken $x$ dependence of the incoherent (and coherent) cross section, for which predictions were made \cite{Mantysaari:2018zdd} using the CGC framework including Jalilian-Marian Iancu  McLerran Weigert Leonidov Kovner (JIMWLK) evolution \cite{Jalilian-Marian:1997jx,Jalilian-Marian:1997gr,Iancu:2000hn,Mueller:2001uk}. We show the ratio of the incoherent to coherent cross section in Fig.\,\ref{fig:jimwlk} (left), where a clear decrease of the ratio with increasing $W$ is visible, in contrast to a calculation in the IPSat model \cite{Kowalski:2003hm,Rezaeian:2012ji,Mantysaari:2018nng}, where the evolution of the spatial structure is neglected. 

Several physics effects lead to this behavior: The JIMWLK evolution will be faster in local regions with small saturation scale than inside the hot spots, which are closer to the saturated regime. This leads to an effective growth of the hot spots and consequently a smoother proton. In addition, as overall $Q_s$ values increase with evolution, the size scale of color charge fluctuations ($\sim 1/Q_s$) decreases, producing more ``color domains" and effectively decreasing geometrical  fluctuations. In the extreme black disk limit, one expects the coherent cross section to dominate, as it receives contributions from the entire proton area, while the incoherent cross section is only sensitive to the edge of the proton.

In Fig.\,\ref{fig:jimwlk} (right) we show the corresponding visualization of the proton as it undergoes JIMWLK evolution, using the same quantity as in Fig.\,\ref{fig:JPsi} (right). Both the growth of the proton and the decrease in the length scale of color charge fluctuations (which are the shortest scales for all rapidities) with increasing rapidity (decreasing $x$) are clearly visible (also see \cite{Schlichting:2014ipa}). The details of this evolution depend on an infrared regulator in the JIMWLK kernel, which is required to avoid violation of the Froissart bound, which based on unitarity arguments puts a constraint on the growth of the inelastic cross section with the collision energy \cite{Froissart:1961ux,Martin:1962rt}. Work on finding  constraints for this regulator, for example in the Gribov-Zwanziger approach for confinement, is ongoing \cite{Gotsman:2020ryd}.

Implementation of a detailed $x$ dependence of the fluctuating nucleon structure in calculations for hadronic and heavy ion collisions is still outstanding. It will be very interesting to see which observables are most sensitive to these effects. Certainly, both collision energy and rapidity dependencies can be used to vary the contributing $x$ values and thereby modify the properties of the projectile and target. For first studies of the implementation of JIMWLK evolution in heavy ion collisions see \cite{Schenke:2016ksl, McDonald:2020oyf}.

It should be noted that an attempt was made within the CGC effective theory at explaining the azimuthal anisotropies (mainly in large systems) from color charge fluctuations of the energy momentum tensor alone, i.e., without including nucleon degrees of freedom at all \cite{Giacalone:2019kgg,Gelis:2019vzt}. The authors start from the connected two-point function of the glasma energy-momentum tensor derived in Ref.~\cite{Albacete:2018bbv}, and by means of an approximation, which neglects all logarithmic corrections to the fluctuations of the glasma energy density, they obtain expressions for the eccentricities of the system which turn out to be compatible with experimental data for realistic choices of the model parameters. However, it has been recently realized \cite{Giacalone:2021} that as soon as the approximations made in the calculation are relaxed, and one makes use of the full McLerran Venugopalan (MV) model expressions of Ref.~\cite{Albacete:2018bbv}, the idea breaks down, as the eccentricities become smaller by one order of magnitude (in agreement with what the numerical calculations within the IP-Glasma model for a smooth nucleus would yield) and can no longer be used to describe the experimental data.


Reversely, another work, which neglects almost everything but the nucleon position fluctuations of the CGC calculation, and dubbed ``Jazma" \cite{Nagle:2018ybc}, reproduces the eccentricities of IP-Glasma in heavy ion collisions well, emphasizing the importance of nucleon position fluctuations, and the limited importance of color charge fluctuations for the initial geometry. However, the model can not produce the full energy momentum tensor of IP-Glasma, which in particular means that it misses the initial state momentum anisotropies that turn out to be important for very small, low multiplicity systems, as will be discussed in detail in Sec.\,\ref{sec:initial}.

In summary, small system collisions have contributed significantly to our improving understanding of the subnucleonic structure. Many observables in a wide variety of experiments, ranging from e+p and e+A collisions, to p+p, p+A (and other small systems), to A+A collisions, prefer a description that includes some level of fluctuating subnucleonic structure, whose details are yet to be understood. Future studies at the EIC, along with new results from small system and heavy ion collisions, are likely to fill in the gaps and provide a much deeper understanding of the fluctuating spatial nucleon and nuclear structure (in addition to the expected advances in understanding the average structure via generalized parton distributions (GPDs) \cite{Diehl:1997bu,Hoodbhoy:1998vm,Belitsky:2000jk,Diehl:2001pm,Belitsky:2001ns}, generalized transverse momentum dependent parton distributions (GTMDs) \cite{Meissner:2009ww,Meissner:2008ay,Lorce:2011dv,Lorce:2013pza}, and Wigner distributions \cite{Ji:2003ak,Belitsky:2003nz}).

\section{More than a fluid -- Initial state momentum anisotropies}
\label{sec:initial}
It was predicted in \cite{Krasnitz:2002ng,Gelis:2008ad,Gelis:2008sz,Dumitru:2008wn,AD_rikenwkshp,Dumitru:2010iy} that multi-gluon production from the CGC leads to long range rapidity correlations that contain azimuthal anisotropies. With the experimental discovery of long range momentum anisotropies in p+p collisions and other small systems \cite{Khachatryan:2010gv,Abelev:2012ola,Aad:2012gla,CMS:2012qk,Adare:2014keg,Adare:2015ctn} many more theoretical calculations, most prominently within the CGC effective theory, were triggered to explain the observations. In these frameworks, no final state effects are necessary to obtain a finite elliptic anisotropy (and in some cases also odd harmonics). We will focus on CGC calculations in this review, but mention that different (but in principle related) frameworks have also been used to try and explain the long range azimuthal anisotropies in small systems without the need for strong final state effects \cite{Levin:2011fb,Gyulassy:2014cfa}. 

CGC calculations are based on solutions of the Yang-Mills equations, which can be obtained numerically, or, under a variety of simplifying assumptions, analytically. Early calculations were based on the glasma graph approximation, which limit the interactions to maximally two-gluon exchanges and uses Gaussian statistics for the initial color charges \cite{Gelis:2008ad,Gelis:2008sz,Dumitru:2008wn,Dumitru:2010mv,Dusling:2012wy,Dusling:2013qoz}.
Keeping Gaussian statistics but resumming multi-gluon exchanges leads to the non-linear Gaussian approximation \cite{McLerran:1998nk,Dominguez:2008aa,Lappi:2015vta}. 
When treated fully numerically \cite{Krasnitz:1998ns,Krasnitz:2002mn,Lappi:2003bi,Schenke:2012wb}, multi-gluon exchanges are included and one has the freedom of using any color charge statistics and realistic spatial distributions \cite{Schenke:2015aqa}. Finally, quantum effects can be included to leading logarithmic order in $\ln(1/x)$ by evolving the initial color charge distributions using e.g.\,the JIMWLK equations \cite{Lappi:2015vha}.

Typically, all of the above calculations find non-zero even harmonics in the long-range two particle correlations, as they have a symmetry in $\vec{k}_1 \rightarrow \vec{k}_2$, $\vec{k}_1 \rightarrow -\vec{k}_1$ and $\vec{k}_2 \rightarrow -\vec{k}_2$, where $\vec{k}_{1,2}$ are the transverse momentum vectors of gluon 1 and 2, respectively. Odd harmonics for gluons generally require an extension of the calculation to including a finite time of Yang-Mills evolution \cite{Lappi:2009xa,Schenke:2015aqa,McLerran:2016snu}, going beyond dilute \emph{and} beyond classical approximations \cite{Kovner:2016jfp,Kovchegov:2018jun} (see a discussion of the dilute approximation in \cite{Schlichting:2019bvy}), or including non-eikonal effects \cite{Agostini:2019avp,Agostini:2019hkj} (also see \cite{Gotsman:2016owk,Gotsman:2016fee} - and comments in \cite{Kovner:2016jfp} - on how odd harmonics could emerge from the Bose-Einstein correlations of identical particles). Quark production within the CGC as in \cite{Dusling:2017dqg,Dusling:2017aot,Mace:2018vwq} does not have the above mentioned symmetries and contains odd harmonics.

The first question one is bound to ask is how such anisotropies emerge within the CGC calculations. There are in fact a variety of sources of anisotropy, both of classical and quantum nature. 

The first emerges from the fact that gluon fields are correlated within a correlation length of $1/Q_s$, which one can loosely interpret as having color field domains of that size. Particles, that scatter (or are produced) from the same domain, are correlated as they feel the same color field. Also density gradients can contribute to anisotropies from the CGC. This is particularly evident in the scattering of a quark-antiquark dipole off a target  \cite{Iancu:2017fzn}. The cross section for dipole target scattering will depend on the dipole orientation if there are significant density gradients in the target. Both effects are purely classical, and in order to generate a correlation, the incoming partons have to be close in the transverse plane, such that they can feel the same local structure of the target. 

Quantum effects also contribute. While the CGC soft gluon state is purely classical, when averaged over the valence color charges, the density matrix is not classical. Evaluation using the weight functional from e.g.~the MV model reveals Bose enhancement effects, an enhanced probability to find two incoming gluons with the same transverse momentum, yielding a similar increase for the produced gluons \cite{Altinoluk:2015uaa}. Finally, there is another quantum effect, namely that of gluonic Hanbury-Brown Twiss (HBT) correlations \cite{Kovchegov:2012nd,Altinoluk:2015uaa}, which appear when one has incoherent emission of identical particles (and is usually used to measure the size of galactic objects using photon correlations \cite{HanburyBrown:1956bqd}). The argument made for realizing an HBT situation is that the scattering randomizes color phases in the projectile on a transverse scale $1/Q_s$, which turns the projectile after the scattering into a collection of sources for incoherent emission.

Having established that and how correlated gluons are produced within the CGC framework, the second question that emerges is of course whether the resulting initial state momentum anisotropies can by themselves explain the observed anisotropies in the experimental data. While early works that studied the associated yield within the near side ridge in p+p and p+Pb collisions (and combined jet as well as glasma contributions to the yield) showed good agreement with the experimental data \cite{Dusling:2012cg,Dusling:2013oia}, more recent calculations, comparing both parton level \cite{Mace:2018vwq} and hadron level results \cite{Greif:2020rhi} in p+Pb collisions to experimental data, underestimate the experimentally observed second and third order harmonics. Hadron level results for azimuthal anisotropies in p+p collisions also underestimate the experimental anisotropies \cite{Schenke:2016lrs}. The difference to the earlier calculations may be that a) It is possible that the harmonics reveal more detail than the associated yield, and b) that including the jet contribution (in the data and the calculation) played a non-negligible role. 

Having to describe hadronization outside of a fluid dynamic description that can make use of an equation of state, complicates the description of the experimental data with purely initial state models (for recent progress see \cite{Greif:2020rhi}). However, so far, the systematics of the experimental data with multiplicity or collision system (e.g. p+Au, d+Au, $^3$He+Au, that were studied at RHIC \cite{PHENIX:2018lia}) could not be reproduced in such frameworks, at least for multiplicities larger than or equal to the minimum bias multiplicity in p+A collisions. This suggests that final state effects are necessary to describe the data, at least for large enough multiplicities. We note that a variety of observables, such as multiparticle cumulants
\cite{Aad:2013fja,Chatrchyan:2013nka,Abelev:2014mda,Khachatryan:2015waa,Yan:2013laa,Skokov:2014tka} or the mass ordering of anisotropy coefficients \cite{Bozek:2013ska,Werner:2013ipa,Schenke:2016lrs,Greif:2020rhi} were suggested to distinguish between initial state and final state pictures. While the CGC calculations struggle with getting the right magnitudes, some qualitative behavior, expected from final state effects, could also be reproduced from the initial state alone \cite{Dumitru:2014yza,Schenke:2016lrs,Greif:2020rhi}.

It is conceivable that as the multiplicity of the collision decreases, final state effects, which rely on the production of a sufficiently strongly interacting system, become less important, and signals of the initial state momentum anisotropy could be observable in the data. We will now discuss an observable that is able to distinguish the origin of the azimuthal anisotropy.

The study of the correlation of the event-by-event mean transverse momentum with the elliptic momentum anisotropy was proposed in \cite{Bozek:2016yoj} to study the correlation of system size and anisotropic flow in small and large systems as an alternative to event shape engineering \cite{Schukraft:2012ah}. This ansatz had the hydrodynamic picture in mind, and showed a remarkable sensitivity to the energy deposition model in p+Pb collisions \cite{Bozek:2016yoj}. 

Specifically, the observable is defined as \cite{Bozek:2016yoj}
\begin{equation}\label{eq:v2PT}
    \rho(v_2^2,[p_T])=\frac{\langle \delta v_2^2 \,\delta [p_T]\rangle}{\sqrt{\langle(\delta v_2^2)^2\rangle\langle(\delta [p_T])^2\rangle}}\,,
\end{equation}
where $[p_T]$ indicates the mean transverse momentum in a single event, $v_2$ is the ($p_T$-integrated) elliptic anisotropy of the particle spectra, and $\delta$ indicates the difference between the single event value and the event average. Typically, $\rho(v_2^2,[p_T])$ is studied at fixed multiplicity \cite{Schenke:2020uqq}. It 
has now been studied experimentally in Pb+Pb and p+Pb collisions by e.g.~the ATLAS Collaboration \cite{Aad:2019fgl,Giacalone:2020dln,Schenke:2020uqq} and is under closer investigation at both RHIC and LHC in a variety of systems, e.g. U+U collisions, as it is sensitive to the quadrupole deformation of the colliding nuclei \cite{Giacalone:2020awm}.

In hydrodynamic frameworks, all initial state models that have so far been used to compute $\rho(v_2^2,[p_T])$ lead to similar qualitative results for heavy ion collisions, when plotted as a function of charged particle multiplicity \cite{Bozek:2016yoj,Giacalone:2020dln,Giacalone:2020awm,Schenke:2020uqq}. These results follow closely predictors for the observable that are based solely on the initial geometry. At small multiplicity (in centrality classes around 70\% or greater for Au+Au and Pb+Pb collisions) the correlator is negative, and turns positive for larger multiplicities. 
This can be understood on the level of the predictors (which are measures of the system size for $[p_T]$ and eccentricity for $v_2$): At small multiplicity, a small area (and by that a large $[p_T]$) is achieved by clustering the participants into a single compact region, which tends to have a smaller eccentricity (resulting in smaller $v_2$). This results in a negative correlation between $[p_T]$ and $v_2$. At larger multiplicity, a smaller area at fixed multiplicity is achieved by fluctuating to a large eccentricity $\varepsilon_2$, as the area of an ellipse is given by $A=\pi[r^2] \sqrt{1-\varepsilon_2^2}$ \cite{Schenke:2020uqq}, which leads to the observed positive correlation between $[p_T]$ and $v_2$.

Interestingly, one expects a positive $\rho(v_2^2,[p_T])$ at small multiplicities, if the anisotropy has initial state momentum correlations as its dominant source. This can be understood as follows. Taking for example the classical source of momentum anisotropy mentioned above, the anisotropy decreases with the number of color domains, which increases with the system size. With $[p_T]$ being well predicted by the initial entropy per area (which is expected also in this initial state scenario, as for fixed multiplicity, which is typically considered, the initial entropy is approximately constant, such that $Q_s$ of the projectile, which drives $[p_T]$, decreases with increasing system size; $Q_s$ of the target is approximately constant), this introduces a positive correlation between $[p_T]$ and $v_2$. This is opposite to the negative correlation produced by geometric effects in small systems.

\begin{figure}[tb]
\centering
\includegraphics[width=0.60\linewidth]{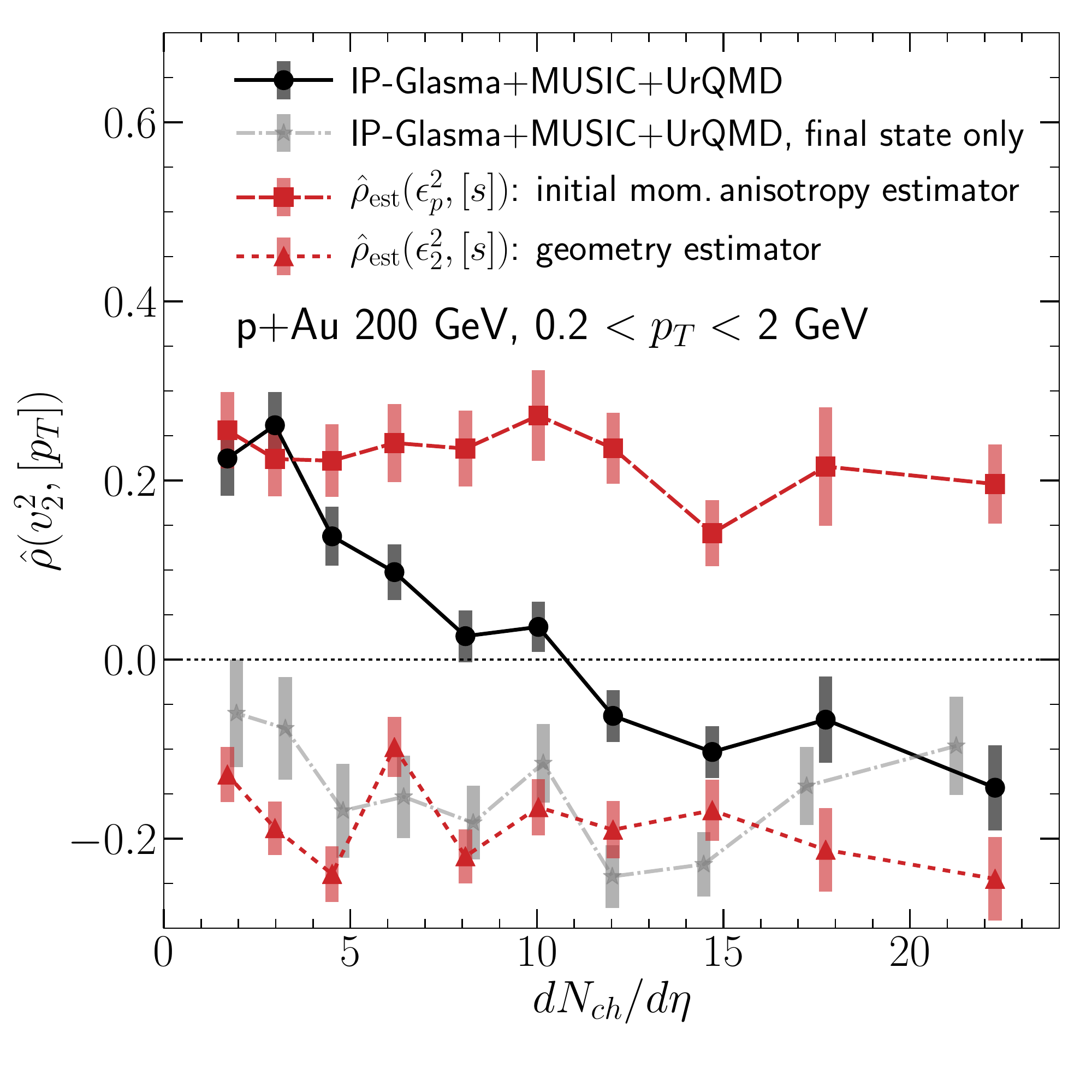}
\caption{The correlation $\hat{\rho}(v_2^2,[p_T])$ of the elliptic flow (squared) with the event-by-event mean transverse momentum $[p_T]$ at fixed multiplicity in $\sqrt{s}=200\,{\rm GeV}$ p+Au collisions. Circles indicate the simulation result including both initial momentum anisotropy and final state evolution. Squares show the expectation for initial momentum anisotropy only, triangles the expectation for final state response to the initial geometry only. Simulation results without initial state momentum anisotropy are shown as stars and follow the expectation for final state effects only. Figure adapted from \cite{Giacalone:2020byk}. \label{fig:rho}}
\end{figure}

Whether these expectations materialize can be tested in a model that contains both initial momentum anisotropies and geometry driven effects, such as the IP-Glasma+\textsc{Music}+UrQMD model, described in detail in \cite{Schenke:2020mbo} (also see \cite{Schenke:2010nt,Schenke:2010rr,Schenke:2011bn} and \cite{Bass:1998ca,Bleicher:1999xi}). Within this model $\rho(v_2^2,[p_T])$ at fixed multiplicity was computed for small systems as a function of charged hadron multiplicity. The result for 200 GeV p+Au collisions is shown in Fig.\,\ref{fig:rho} as solid circles (connected with solid lines). One can see that the sign of $\rho(v_2^2,[p_T])$ changes from positive to negative as one increases the multiplicity. In fact, studying the geometric predictor (triangles, dotted line) and the predictor from the initial momentum anisotropy (squares, dashed line), one sees that the final result moves from one predictor to the other as the multiplicity changes. This indicates that within the simulation the initial state momentum anisotropy begins to dominate as one decreases the multiplicity to approximately 10 charged hadrons per unit rapidity, as this is where   $\rho(v_2^2,[p_T])$ changes sign. A simulation with the initial momentum anisotropy turned off (stars, dash-dotted line) agrees well with the geometric predictor as expected. Similar results were found for d+Au collisions at RHIC and p+Pb collisions at the LHC. Experimental confirmation of the sign change (when eliminating any kind of non-flow correlations, which is difficult at such small multiplicities \cite{Zhang:2021phk}) will be a strong indication of the existence and importance of initial state momentum anisotropies, as predicted by the CGC effective theory.

Other complex observables, such as normalized symmetric cumulants \cite{Bilandzic:2013kga}, may also have the potential to distinguish the origin of the azimuthal momentum anisotropy in small systems. More detailed phenomenological studies together with sophisticated experimental analyses at low multiplicity will likely shed more light on this in the near future. If one can confirm the picture shown in Fig.\,\ref{fig:rho}, which indicates that small systems do behave like fluids, but that at low multiplicities we have access to multi-gluon correlations governed by QCD, it would in fact be the most exciting outcome.

\section{Conclusions and Outlook}
Small system (p+p, p/d/$^3$He+A) collisions at RHIC and LHC have driven theory developments towards understanding conditions for the applicability of (relativistic) fluid dynamics as well as particle production and properties of nuclei and hadrons at high energy. We have reviewed three major developments. 

First, our understanding of the applicability of hydrodynamics in systems away from equilibrium, using investigations of hydrodynamic and non-hydrodynamic modes, as well as attractor behavior in kinetic theory or holographic theories, has significantly improved over the last several years. 
Second, the role of subnucleon fluctuations has been increasingly recognized. The substructure of the proton seems to have important implications for both e+p as well as p+p and p+A collisions. This connects a wide range of experiments, and new calculations modeling a hot spot substructure within the CGC EFT paint a consistent picture among them. 
Finally, the CGC predicts azimuthal correlations between produced particles even in the absence of final state effects. Attempts to systematically describe experimental data using only these initial state effects have failed. Nevertheless, it is conceivable that both initial and final state effects affect the observable azimuthal anisotropies of produced charged particles in small system collisions, with the role of the initial state contribution increasing with decreasing system size (or particle multiplicity). We have presented recent developments in identifying observables that could be used to distinguish the two sources of anisotropies. It would be most exciting if one could show that both fluid behavior is present, i.e., the world's smallest fluid is produced in p+A or p+p collisions, but access to complex gluon correlations, directly driven by QCD, is also possible.

It should be noted that interesting developments are also ongoing within the high energy physics community to extend event generators such as PYTHIA \cite{Sjostrand:2014zea} to be able to reproduce the momentum anisotropies observed in p+p collisions. This involves for example the inclusion of a dynamically generated transverse pressure, produced by the excess energy from overlapping strings \cite{Bierlich:2016vgw} in the rope hadronization picture \cite{Bierlich:2014xba}, which has similar effects as the final state evolution discussed here.

To make progress, more effort needs to be geared towards understanding the full three dimensional dynamics of small system collisions, which will contribute an important additional handle. Many current calculations assume boost invariance but, particularly at RHIC energies, the variation of observables in the rapidity direction (along the beam line) is significant \cite{Aidala:2017pup}. Besides hydrodynamic and transport approaches \cite{Bozek:2014cya,Bzdak:2014dia,Koop:2015wea,Shen:2017bsr}, also Color Glass Condensate based calculations are being extended to three spatial dimensions \cite{Schenke:2016ksl,McDonald:2020oyf,Schlichting:2020wrv}. Understanding the rapidity dependence of initial momentum anisotropies and final state generated anisotropies will be of extraordinary value for identifying the dominant source of the experimentally observed anisotropy, especially when considering the planned experimental improvements, such as the recently approved forward upgrade program at STAR. In particular, having information on the expected rapidity dependence of e.g. initial momentum anisotropies and geometry driven anisotropies, will be helpful in distinguishing them from non-flow (short range correlations from e.g. resonance decays and jets) in the experimental data.

One big uncertainty is the early time evolution, which in small systems can take up a significant amount of the entire lifetime. To make progress, non-equilibrium stages (described by e.g.~effective kinetic theory) in simulations that couple initial conditions to hydrodynamics need to be included also in small system studies. Improvements of the non-equilibrium evolution, such as extensions to non-conformal theories and better handling of large gradients, are also required. 

We have not addressed particle production at high transverse momentum here, but the question of jet quenching and azimuthal anisotropies at high $p_T$ are also very important to gain a complete understanding of small systems \cite{Nagle:2018nvi}. While a significant suppression of particles at high transverse momentum in A+A collisions relative to p+p collisions is observed experimentally, p/d+A collisions show almost no modification \cite{Adler:2003ii,Adams:2003im,Arsene:2003yk,Back:2003ns,Khachatryan:2016odn} (and it may not be expected theoretically \cite{Tywoniuk:2014hta,Shen:2016egw}). On the other hand, a reasonably strong elliptic anisotropy is observed at high $p_T$ in p+Pb collisions, whose origin is explained by directionally dependent jet quenching in A+A collisions, but appears mysterious in p+Pb collisions that show no jet quenching. 
Also, as centrality selection is problematic in small systems, which complicates the definition of the nuclear modification factor \cite{ATLAS:2014cpa}, the proposed study of O+O collisions \cite{Huang:2019tgz} at LHC 
and RHIC, could prove useful in exploring jet quenching in small systems. Comparing O+O to p+Pb collisions at a similar number of produced particles could yield differences in jet quenching as the initial system sizes can differ considerably. The study of O+O along with p+A is also argued to help with separating initial momentum anisotropy from final state effects \cite{Huang:2019tgz}.

Finally, measurements of electromagnetic probes could be used to further support the interpretation of the formation of a strongly interacting medium in small system collisions \cite{Shen:2015qba}. Additional photon and dilepton radiation from the medium should be visible, as it is in heavy ion collisions. While the existing experimental data from RHIC \cite{Adare:2012vn} is compatible with the expected enhancement over the expectations from p+p collisions, its current precision does not allow for strong conclusions to be drawn.

In closing, we believe it is fair to say that small system collisions at RHIC and LHC have driven tremendous theoretical progress that is rippling through fields far beyond that of heavy ion physics. As many open questions remain and work is ongoing both on the experimental and theory side, we are likely to see this progress continue well into the future.

\section{Acknowledgments}
B.P.S. thanks Charles Gale, Giuliano Giacalone, Jiangyong Jia, Heikki M\"antysaari, Aleksas Mazeliauskas, Jean-Fran\c{c}ois Paquet, S\"oren Schlichting, Prithwish Tribedy, Chun Shen, and Raju Venugopalan for useful discussions. Special thanks go to Charles Gale, Giuliano Giacalone, S\"oren Schlichting, Chun Shen, and Michael Strickland for valuable comments on an early version of the manuscript. 
B.P.S. is supported under DOE Contract No. DE-SC0012704. 
\vspace{1cm}

\bibliographystyle{iopart-num}
\bibliography{references}

\providecommand{\newblock}{}
\begin{thebibliography}{100}
\expandafter\ifx\csname url\endcsname\relax
  \def\url#1{{\tt #1}}\fi
\expandafter\ifx\csname urlprefix\endcsname\relax\def\urlprefix{URL }\fi
\providecommand{\eprint}[2][]{\url{#2}}

\bibitem{Arsene:2004fa}
Arsene I {\em et~al.\/} (BRAHMS) 2005 {\em Nucl. Phys. A\/} {\bf 757} 1--27
  (\textit{Preprint} \eprint{nucl-ex/0410020})

\bibitem{Back:2004je}
Back B~B {\em et~al.\/} (PHOBOS) 2005 {\em Nucl. Phys. A\/} {\bf 757} 28--101
  (\textit{Preprint} \eprint{nucl-ex/0410022})

\bibitem{Adams:2005dq}
Adams J {\em et~al.\/} (STAR) 2005 {\em Nucl. Phys. A\/} {\bf 757} 102--183
  (\textit{Preprint} \eprint{nucl-ex/0501009})

\bibitem{Adcox:2004mh}
Adcox K {\em et~al.\/} (PHENIX) 2005 {\em Nucl. Phys. A\/} {\bf 757} 184--283
  (\textit{Preprint} \eprint{nucl-ex/0410003})

\bibitem{ALICE:2011ab}
Aamodt K {\em et~al.\/} (ALICE) 2011 {\em Phys. Rev. Lett.\/} {\bf 107} 032301
  (\textit{Preprint} \eprint{1105.3865})

\bibitem{ATLAS:2011ah}
Aad G {\em et~al.\/} (ATLAS) 2012 {\em Phys. Lett. B\/} {\bf 707} 330--348
  (\textit{Preprint} \eprint{1108.6018})

\bibitem{ATLAS:2012at}
Aad G {\em et~al.\/} (ATLAS) 2012 {\em Phys. Rev. C\/} {\bf 86} 014907
  (\textit{Preprint} \eprint{1203.3087})

\bibitem{Chatrchyan:2012wg}
Chatrchyan S {\em et~al.\/} (CMS) 2012 {\em Eur. Phys. J. C\/} {\bf 72} 2012
  (\textit{Preprint} \eprint{1201.3158})

\bibitem{Heinz:2013th}
Heinz U and Snellings R 2013 {\em Ann. Rev. Nucl. Part. Sci.\/} {\bf 63}
  123--151 (\textit{Preprint} \eprint{1301.2826})

\bibitem{Gale:2013da}
Gale C, Jeon S and Schenke B 2013 {\em Int. J. Mod. Phys. A\/} {\bf 28} 1340011
  (\textit{Preprint} \eprint{1301.5893})

\bibitem{Gajdosova:2020nvb}
K\v{r}\'\i{}\v{z}kov\'a~Gajdo\v{s}ov\'a K 2021 {\em Nucl. Phys. A\/} {\bf 1005}
  121802 (\textit{Preprint} \eprint{2007.12529})

\bibitem{Khachatryan:2010gv}
Khachatryan V {\em et~al.\/} (CMS) 2010 {\em JHEP\/} {\bf 09} 091
  (\textit{Preprint} \eprint{1009.4122})

\bibitem{Abelev:2012ola}
Abelev B {\em et~al.\/} (ALICE) 2013 {\em Phys. Lett. B\/} {\bf 719} 29--41
  (\textit{Preprint} \eprint{1212.2001})

\bibitem{Aad:2012gla}
Aad G {\em et~al.\/} (ATLAS) 2013 {\em Phys. Rev. Lett.\/} {\bf 110} 182302
  (\textit{Preprint} \eprint{1212.5198})

\bibitem{CMS:2012qk}
Chatrchyan S {\em et~al.\/} (CMS) 2013 {\em Phys. Lett. B\/} {\bf 718} 795--814
  (\textit{Preprint} \eprint{1210.5482})

\bibitem{Aidala:2016vgl}
Aidala C {\em et~al.\/} (PHENIX) 2017 {\em Phys. Rev. C\/} {\bf 95} 034910
  (\textit{Preprint} \eprint{1609.02894})

\bibitem{Adam:2019woz}
Adam J {\em et~al.\/} (STAR) 2019 {\em Phys. Rev. Lett.\/} {\bf 122} 172301
  (\textit{Preprint} \eprint{1901.08155})

\bibitem{Adare:2013piz}
Adare A {\em et~al.\/} (PHENIX) 2013 {\em Phys. Rev. Lett.\/} {\bf 111} 212301
  (\textit{Preprint} \eprint{1303.1794})

\bibitem{Adare:2014keg}
Adare A {\em et~al.\/} (PHENIX) 2015 {\em Phys. Rev. Lett.\/} {\bf 114} 192301
  (\textit{Preprint} \eprint{1404.7461})

\bibitem{Adamczyk:2014fcx}
Adamczyk L {\em et~al.\/} (STAR) 2015 {\em Phys. Lett. B\/} {\bf 743} 333--339
  (\textit{Preprint} \eprint{1412.8437})

\bibitem{Adare:2015ctn}
Adare A {\em et~al.\/} (PHENIX) 2015 {\em Phys. Rev. Lett.\/} {\bf 115} 142301
  (\textit{Preprint} \eprint{1507.06273})

\bibitem{Lacey:2020ime}
Lacey R~A (STAR) 2021 {\em Nucl. Phys. A\/} {\bf 1005} 122041
  (\textit{Preprint} \eprint{2002.11889})

\bibitem{Dusling:2015gta}
Dusling K, Li W and Schenke B 2016 {\em Int. J. Mod. Phys. E\/} {\bf 25}
  1630002 (\textit{Preprint} \eprint{1509.07939})

\bibitem{Loizides:2016tew}
Loizides C 2016 {\em Nucl. Phys. A\/} {\bf 956} 200--207 (\textit{Preprint}
  \eprint{1602.09138})

\bibitem{Schlichting:2016sqo}
Schlichting S and Tribedy P 2016 {\em Adv. High Energy Phys.\/} {\bf 2016}
  8460349 (\textit{Preprint} \eprint{1611.00329})

\bibitem{Nagle:2018nvi}
Nagle J~L and Zajc W~A 2018 {\em Ann. Rev. Nucl. Part. Sci.\/} {\bf 68}
  211--235 (\textit{Preprint} \eprint{1801.03477})

\bibitem{Marti:1999wi}
Marti J~M and Mueller E 1999 {\em Living Rev. Rel.\/} {\bf 2} 3
  (\textit{Preprint} \eprint{astro-ph/9906333})

\bibitem{Font:2000pp}
Font J~A 2000 {\em Living Rev. Rel.\/} {\bf 3} 2 (\textit{Preprint}
  \eprint{gr-qc/0003101})

\bibitem{Accardi:2012qut}
Accardi A {\em et~al.\/} 2016 {\em Eur. Phys. J. A\/} {\bf 52} 268
  (\textit{Preprint} \eprint{1212.1701})

\bibitem{Bozek:2011if}
Bozek P 2012 {\em Phys. Rev. C\/} {\bf 85} 014911 (\textit{Preprint}
  \eprint{1112.0915})

\bibitem{Bozek:2012gr}
Bozek P and Broniowski W 2013 {\em Phys. Lett. B\/} {\bf 718} 1557--1561
  (\textit{Preprint} \eprint{1211.0845})

\bibitem{Bozek:2013df}
Bozek P and Broniowski W 2013 {\em Phys. Lett. B\/} {\bf 720} 250--253
  (\textit{Preprint} \eprint{1301.3314})

\bibitem{Bozek:2013uha}
Bozek P and Broniowski W 2013 {\em Phys. Rev. C\/} {\bf 88} 014903
  (\textit{Preprint} \eprint{1304.3044})

\bibitem{Bozek:2013ska}
Bozek P, Broniowski W and Torrieri G 2013 {\em Phys. Rev. Lett.\/} {\bf 111}
  172303 (\textit{Preprint} \eprint{1307.5060})

\bibitem{Bzdak:2013zma}
Bzdak A, Schenke B, Tribedy P and Venugopalan R 2013 {\em Phys. Rev. C\/} {\bf
  87} 064906 (\textit{Preprint} \eprint{1304.3403})

\bibitem{Qin:2013bha}
Qin G~Y and M\"uller B 2014 {\em Phys. Rev. C\/} {\bf 89} 044902
  (\textit{Preprint} \eprint{1306.3439})

\bibitem{Werner:2013ipa}
Werner K, Bleicher M, Guiot B, Karpenko I and Pierog T 2014 {\em Phys. Rev.
  Lett.\/} {\bf 112} 232301 (\textit{Preprint} \eprint{1307.4379})

\bibitem{Kozlov:2014fqa}
Kozlov I, Luzum M, Denicol G, Jeon S and Gale C 2014  (\textit{Preprint}
  \eprint{1405.3976})

\bibitem{Romatschke:2015gxa}
Romatschke P 2015 {\em Eur. Phys. J. C\/} {\bf 75} 305 (\textit{Preprint}
  \eprint{1502.04745})

\bibitem{Shen:2016zpp}
Shen C, Paquet J~F, Denicol G~S, Jeon S and Gale C 2017 {\em Phys. Rev. C\/}
  {\bf 95} 014906 (\textit{Preprint} \eprint{1609.02590})

\bibitem{Weller:2017tsr}
Weller R~D and Romatschke P 2017 {\em Phys. Lett. B\/} {\bf 774} 351--356
  (\textit{Preprint} \eprint{1701.07145})

\bibitem{Arnold:2004ti}
Arnold P~B, Lenaghan J, Moore G~D and Yaffe L~G 2005 {\em Phys. Rev. Lett.\/}
  {\bf 94} 072302 (\textit{Preprint} \eprint{nucl-th/0409068})

\bibitem{Romatschke:2016hle}
Romatschke P 2017 {\em Eur. Phys. J. C\/} {\bf 77} 21 (\textit{Preprint}
  \eprint{1609.02820})

\bibitem{Chesler:2009cy}
Chesler P~M and Yaffe L~G 2010 {\em Phys. Rev. D\/} {\bf 82} 026006
  (\textit{Preprint} \eprint{0906.4426})

\bibitem{Heller:2011ju}
Heller M~P, Janik R~A and Witaszczyk P 2012 {\em Phys. Rev. Lett.\/} {\bf 108}
  201602 (\textit{Preprint} \eprint{1103.3452})

\bibitem{Casalderrey-Solana:2013aba}
Casalderrey-Solana J, Heller M~P, Mateos D and van~der Schee W 2013 {\em Phys.
  Rev. Lett.\/} {\bf 111} 181601 (\textit{Preprint} \eprint{1305.4919})

\bibitem{Kurkela:2015qoa}
Kurkela A and Zhu Y 2015 {\em Phys. Rev. Lett.\/} {\bf 115} 182301
  (\textit{Preprint} \eprint{1506.06647})

\bibitem{Keegan:2015avk}
Keegan L, Kurkela A, Romatschke P, van~der Schee W and Zhu Y 2016 {\em JHEP\/}
  {\bf 04} 031 (\textit{Preprint} \eprint{1512.05347})

\bibitem{Heller:2013fn}
Heller M~P, Janik R~A and Witaszczyk P 2013 {\em Phys. Rev. Lett.\/} {\bf 110}
  211602 (\textit{Preprint} \eprint{1302.0697})

\bibitem{Heller:2015dha}
Heller M~P and Spalinski M 2015 {\em Phys. Rev. Lett.\/} {\bf 115} 072501
  (\textit{Preprint} \eprint{1503.07514})

\bibitem{Buchel:2016cbj}
Buchel A, Heller M~P and Noronha J 2016 {\em Phys. Rev. D\/} {\bf 94} 106011
  (\textit{Preprint} \eprint{1603.05344})

\bibitem{Denicol:2016bjh}
Denicol G~S and Noronha J 2016  (\textit{Preprint} \eprint{1608.07869})

\bibitem{Heller:2016rtz}
Heller M~P, Kurkela A, Spali\'nski M and Svensson V 2018 {\em Phys. Rev. D\/}
  {\bf 97} 091503 (\textit{Preprint} \eprint{1609.04803})

\bibitem{Florkowski:2017olj}
Florkowski W, Heller M~P and Spalinski M 2018 {\em Rept. Prog. Phys.\/} {\bf
  81} 046001 (\textit{Preprint} \eprint{1707.02282})

\bibitem{Romatschke:2017vte}
Romatschke P 2018 {\em Phys. Rev. Lett.\/} {\bf 120} 012301 (\textit{Preprint}
  \eprint{1704.08699})

\bibitem{Bemfica:2017wps}
Bemfica F~S, Disconzi M~M and Noronha J 2018 {\em Phys. Rev. D\/} {\bf 98}
  104064 (\textit{Preprint} \eprint{1708.06255})

\bibitem{Spalinski:2017mel}
Spali\'nski M 2018 {\em Phys. Lett. B\/} {\bf 776} 468--472 (\textit{Preprint}
  \eprint{1708.01921})

\bibitem{Romatschke:2017acs}
Romatschke P 2017 {\em JHEP\/} {\bf 12} 079 (\textit{Preprint}
  \eprint{1710.03234})

\bibitem{Behtash:2017wqg}
Behtash A, Cruz-Camacho C~N and Martinez M 2018 {\em Phys. Rev. D\/} {\bf 97}
  044041 (\textit{Preprint} \eprint{1711.01745})

\bibitem{Florkowski:2017jnz}
Florkowski W, Maksymiuk E and Ryblewski R 2018 {\em Phys. Rev. C\/} {\bf 97}
  024915 (\textit{Preprint} \eprint{1710.07095})

\bibitem{Florkowski:2017ovw}
Florkowski W, Maksymiuk E and Ryblewski R 2018 {\em Phys. Rev. C\/} {\bf 97}
  014904 (\textit{Preprint} \eprint{1711.03872})

\bibitem{Strickland:2017kux}
Strickland M, Noronha J and Denicol G 2018 {\em Phys. Rev. D\/} {\bf 97} 036020
  (\textit{Preprint} \eprint{1709.06644})

\bibitem{Almaalol:2018ynx}
Almaalol D and Strickland M 2018 {\em Phys. Rev. C\/} {\bf 97} 044911
  (\textit{Preprint} \eprint{1801.10173})

\bibitem{Denicol:2018pak}
Denicol G~S and Noronha J 2019 {\em Phys. Rev. D\/} {\bf 99} 116004
  (\textit{Preprint} \eprint{1804.04771})

\bibitem{Behtash:2018moe}
Behtash A, Cruz-Camacho C~N, Kamata S and Martinez M 2019 {\em Phys. Lett. B\/}
  {\bf 797} 134914 (\textit{Preprint} \eprint{1805.07881})

\bibitem{Strickland:2018ayk}
Strickland M 2018 {\em JHEP\/} {\bf 12} 128 (\textit{Preprint}
  \eprint{1809.01200})

\bibitem{Heller:2018qvh}
Heller M~P and Svensson V 2018 {\em Phys. Rev. D\/} {\bf 98} 054016
  (\textit{Preprint} \eprint{1802.08225})

\bibitem{Mazeliauskas:2018yef}
Mazeliauskas A and Berges J 2019 {\em Phys. Rev. Lett.\/} {\bf 122} 122301
  (\textit{Preprint} \eprint{1810.10554})

\bibitem{Behtash:2019qtk}
Behtash A, Kamata S, Martinez M and Shi H 2020 {\em JHEP\/} {\bf 07} 226
  (\textit{Preprint} \eprint{1911.06406})

\bibitem{Strickland:2019hff}
Strickland M and Tantary U 2019 {\em JHEP\/} {\bf 10} 069 (\textit{Preprint}
  \eprint{1903.03145})

\bibitem{Jaiswal:2019cju}
Jaiswal S, Chattopadhyay C, Jaiswal A, Pal S and Heinz U 2019 {\em Phys. Rev.
  C\/} {\bf 100} 034901 (\textit{Preprint} \eprint{1907.07965})

\bibitem{Kurkela:2019set}
Kurkela A, van~der Schee W, Wiedemann U~A and Wu B 2020 {\em Phys. Rev.
  Lett.\/} {\bf 124} 102301 (\textit{Preprint} \eprint{1907.08101})

\bibitem{Chattopadhyay:2019jqj}
Chattopadhyay C and Heinz U~W 2020 {\em Phys. Lett. B\/} {\bf 801} 135158
  (\textit{Preprint} \eprint{1911.07765})

\bibitem{Brewer:2019oha}
Brewer J, Yan L and Yin Y 2019  (\textit{Preprint} \eprint{1910.00021})

\bibitem{Kurkela:2019kip}
Kurkela A, Wiedemann U~A and Wu B 2019 {\em Eur. Phys. J. C\/} {\bf 79} 965
  (\textit{Preprint} \eprint{1905.05139})

\bibitem{Almaalol:2020rnu}
Almaalol D, Kurkela A and Strickland M 2020 {\em Phys. Rev. Lett.\/} {\bf 125}
  122302 (\textit{Preprint} \eprint{2004.05195})

\bibitem{Muller:1967zza}
Muller I 1967 {\em Z. Phys.\/} {\bf 198} 329--344

\bibitem{Israel:1976tn}
Israel W 1976 {\em Annals Phys.\/} {\bf 100} 310--331

\bibitem{Israel:1979wp}
Israel W and Stewart J~M 1979 {\em Annals Phys.\/} {\bf 118} 341--372

\bibitem{Muller:1999in}
Muller I 1999 {\em Living Rev. Rel.\/} {\bf 2} 1

\bibitem{Baier:2007ix}
Baier R, Romatschke P, Son D~T, Starinets A~O and Stephanov M~A 2008 {\em
  JHEP\/} {\bf 04} 100 (\textit{Preprint} \eprint{0712.2451})

\bibitem{Denicol:2012cn}
Denicol G~S, Niemi H, Molnar E and Rischke D~H 2012 {\em Phys. Rev. D\/} {\bf
  85} 114047 [Erratum: Phys.Rev.D 91, 039902 (2015)] (\textit{Preprint}
  \eprint{1202.4551})

\bibitem{Florkowski:2010cf}
Florkowski W and Ryblewski R 2011 {\em Phys. Rev. C\/} {\bf 83} 034907
  (\textit{Preprint} \eprint{1007.0130})

\bibitem{Martinez:2010sc}
Martinez M and Strickland M 2010 {\em Nucl. Phys. A\/} {\bf 848} 183--197
  (\textit{Preprint} \eprint{1007.0889})

\bibitem{Ryblewski:2010bs}
Ryblewski R and Florkowski W 2011 {\em J. Phys. G\/} {\bf 38} 015104
  (\textit{Preprint} \eprint{1007.4662})

\bibitem{Florkowski:2013lza}
Florkowski W, Ryblewski R and Strickland M 2013 {\em Nucl. Phys. A\/} {\bf 916}
  249--259 (\textit{Preprint} \eprint{1304.0665})

\bibitem{Nopoush:2014pfa}
Nopoush M, Ryblewski R and Strickland M 2014 {\em Phys. Rev. C\/} {\bf 90}
  014908 (\textit{Preprint} \eprint{1405.1355})

\bibitem{Nopoush:2014qba}
Nopoush M, Ryblewski R and Strickland M 2015 {\em Phys. Rev. D\/} {\bf 91}
  045007 (\textit{Preprint} \eprint{1410.6790})

\bibitem{McNelis:2018jho}
McNelis M, Bazow D and Heinz U 2018 {\em Phys. Rev. C\/} {\bf 97} 054912
  (\textit{Preprint} \eprint{1803.01810})

\bibitem{Strickland:2014pga}
Strickland M 2014 {\em Acta Phys. Polon. B\/} {\bf 45} 2355--2394
  (\textit{Preprint} \eprint{1410.5786})

\bibitem{Alqahtani:2017mhy}
Alqahtani M, Nopoush M and Strickland M 2018 {\em Prog. Part. Nucl. Phys.\/}
  {\bf 101} 204--248 (\textit{Preprint} \eprint{1712.03282})

\bibitem{Martinez:2017ibh}
Martinez M, McNelis M and Heinz U 2017 {\em Phys. Rev. C\/} {\bf 95} 054907
  (\textit{Preprint} \eprint{1703.10955})

\bibitem{Chattopadhyay:2018apf}
Chattopadhyay C, Heinz U, Pal S and Vujanovic G 2018 {\em Phys. Rev. C\/} {\bf
  97} 064909 (\textit{Preprint} \eprint{1801.07755})

\bibitem{Berges:2013fga}
Berges J, Boguslavski K, Schlichting S and Venugopalan R 2014 {\em Phys. Rev.
  D\/} {\bf 89} 114007 (\textit{Preprint} \eprint{1311.3005})

\bibitem{Baier:2000sb}
Baier R, Mueller A~H, Schiff D and Son D~T 2001 {\em Phys. Lett. B\/} {\bf 502}
  51--58 (\textit{Preprint} \eprint{hep-ph/0009237})

\bibitem{Arnold:2002zm}
Arnold P~B, Moore G~D and Yaffe L~G 2003 {\em JHEP\/} {\bf 01} 030
  (\textit{Preprint} \eprint{hep-ph/0209353})

\bibitem{Kurkela:2018vqr}
Kurkela A, Mazeliauskas A, Paquet J~F, Schlichting S and Teaney D 2019 {\em
  Phys. Rev. C\/} {\bf 99} 034910 (\textit{Preprint} \eprint{1805.00961})

\bibitem{Kurkela:2018wud}
Kurkela A, Mazeliauskas A, Paquet J~F, Schlichting S and Teaney D 2019 {\em
  Phys. Rev. Lett.\/} {\bf 122} 122302 (\textit{Preprint} \eprint{1805.01604})

\bibitem{Schenke:2012wb}
Schenke B, Tribedy P and Venugopalan R 2012 {\em Phys. Rev. Lett.\/} {\bf 108}
  252301 (\textit{Preprint} \eprint{1202.6646})

\bibitem{Schenke:2012hg}
Schenke B, Tribedy P and Venugopalan R 2012 {\em Phys. Rev. C\/} {\bf 86}
  034908 (\textit{Preprint} \eprint{1206.6805})

\bibitem{Kurkela:2016vts}
Kurkela A 2016 {\em Nucl. Phys. A\/} {\bf 956} 136--143 (\textit{Preprint}
  \eprint{1601.03283})

\bibitem{Kurkela:2020wwb}
Kurkela A, Taghavi S~F, Wiedemann U~A and Wu B 2020 {\em Phys. Lett. B\/} {\bf
  811} 135901 (\textit{Preprint} \eprint{2007.06851})

\bibitem{Giacalone:2019ldn}
Giacalone G, Mazeliauskas A and Schlichting S 2019 {\em Phys. Rev. Lett.\/}
  {\bf 123} 262301 (\textit{Preprint} \eprint{1908.02866})

\bibitem{Kamata:2020mka}
Kamata S, Martinez M, Plaschke P, Ochsenfeld S and Schlichting S 2020 {\em
  Phys. Rev. D\/} {\bf 102} 056003 (\textit{Preprint} \eprint{2004.06751})

\bibitem{Kasmaei:2019ofu}
Kasmaei B~S and Strickland M 2020 {\em Phys. Rev. D\/} {\bf 102} 014037
  (\textit{Preprint} \eprint{1911.03370})

\bibitem{Gale:2020dum}
Gale C, Paquet J~F, Schenke B and Shen C 2020 {Event-plane decorrelation of
  photons produced in the early stage of heavy-ion collisions} {\em {10th
  International Conference on Hard and Electromagnetic Probes of High-Energy
  Nuclear Collisions}: {Hard Probes 2020~}\/} (\textit{Preprint}
  \eprint{2009.07841})

\bibitem{Churchill:2020uvk}
Churchill J, Yan L, Jeon S and Gale C 2021 {\em Phys. Rev. C\/} {\bf 103}
  024904 (\textit{Preprint} \eprint{2008.02902})

\bibitem{Kurkela:2018xxd}
Kurkela A and Mazeliauskas A 2019 {\em Phys. Rev. Lett.\/} {\bf 122} 142301
  (\textit{Preprint} \eprint{1811.03040})

\bibitem{Kurkela:2018oqw}
Kurkela A and Mazeliauskas A 2019 {\em Phys. Rev. D\/} {\bf 99} 054018
  (\textit{Preprint} \eprint{1811.03068})

\bibitem{Du:2020dvp}
Du X and Schlichting S 2020  (\textit{Preprint} \eprint{2012.09079})

\bibitem{Du:2020zqg}
Du X and Schlichting S 2020  (\textit{Preprint} \eprint{2012.09068})

\bibitem{Gale:2020xlg}
Gale C, Paquet J~F, Schenke B and Shen C 2021 {\em Nucl. Phys. A\/} {\bf 1005}
  121863 (\textit{Preprint} \eprint{2002.05191})

\bibitem{NunesdaSilva:2020bfs}
Nunes~da Silva T, Chinellato D, Hippert M, Serenone W, Takahashi J, Denicol
  G~S, Luzum M and Noronha J 2020  (\textit{Preprint} \eprint{2006.02324})

\bibitem{Duez:2004nf}
Duez M~D, Liu Y~T, Shapiro S~L and Stephens B~C 2004 {\em Phys. Rev. D\/} {\bf
  69} 104030 (\textit{Preprint} \eprint{astro-ph/0402502})

\bibitem{Shibata:2017xht}
Shibata M and Kiuchi K 2017 {\em Phys. Rev. D\/} {\bf 95} 123003
  (\textit{Preprint} \eprint{1705.06142})

\bibitem{Most:2018eaw}
Most E~R, Papenfort L~J, Dexheimer V, Hanauske M, Schramm S, St\"ocker H and
  Rezzolla L 2019 {\em Phys. Rev. Lett.\/} {\bf 122} 061101 (\textit{Preprint}
  \eprint{1807.03684})

\bibitem{Alford:2019kdw}
Alford M, Harutyunyan A and Sedrakian A 2019 {\em Phys. Rev. D\/} {\bf 100}
  103021 (\textit{Preprint} \eprint{1907.04192})

\bibitem{Sachdev:2010ch}
Sachdev S 2011 {\em Lect. Notes Phys.\/} {\bf 828} 273--311 (\textit{Preprint}
  \eprint{1002.2947})

\bibitem{Nastase:2017cxp}
Nastase H 2017 {\em {String Theory Methods for Condensed Matter Physics}\/}
  (Cambridge University Press) ISBN 978-1-316-85304-7, 978-1-107-18038-3

\bibitem{Schlichting:2019abc}
Schlichting S and Teaney D 2019 {\em Ann. Rev. Nucl. Part. Sci.\/} {\bf 69}
  447--476 (\textit{Preprint} \eprint{1908.02113})

\bibitem{Berges:2020fwq}
Berges J, Heller M~P, Mazeliauskas A and Venugopalan R 2020  (\textit{Preprint}
  \eprint{2005.12299})

\bibitem{Alver:2010gr}
Alver B and Roland G 2010 {\em Phys. Rev. C\/} {\bf 81} 054905 [Erratum:
  Phys.Rev.C 82, 039903 (2010)] (\textit{Preprint} \eprint{1003.0194})

\bibitem{Borghini:2000sa}
Borghini N, Dinh P~M and Ollitrault J~Y 2001 {\em Phys. Rev. C\/} {\bf 63}
  054906 (\textit{Preprint} \eprint{nucl-th/0007063})

\bibitem{Borghini:2001vi}
Borghini N, Dinh P~M and Ollitrault J~Y 2001 {\em Phys. Rev. C\/} {\bf 64}
  054901 (\textit{Preprint} \eprint{nucl-th/0105040})

\bibitem{Bilandzic:2013kga}
Bilandzic A, Christensen C~H, Gulbrandsen K, Hansen A and Zhou Y 2014 {\em
  Phys. Rev. C\/} {\bf 89} 064904 (\textit{Preprint} \eprint{1312.3572})

\bibitem{Miller:2007ri}
Miller M~L, Reygers K, Sanders S~J and Steinberg P 2007 {\em Ann. Rev. Nucl.
  Part. Sci.\/} {\bf 57} 205--243 (\textit{Preprint} \eprint{nucl-ex/0701025})

\bibitem{dEnterria:2020dwq}
d'Enterria D and Loizides C 2020  (\textit{Preprint} \eprint{2011.14909})

\bibitem{Giacalone:2017uqx}
Giacalone G, Noronha-Hostler J and Ollitrault J~Y 2017 {\em Phys. Rev. C\/}
  {\bf 95} 054910 (\textit{Preprint} \eprint{1702.01730})

\bibitem{Moreland:2014oya}
Moreland J~S, Bernhard J~E and Bass S~A 2015 {\em Phys. Rev. C\/} {\bf 92}
  011901 (\textit{Preprint} \eprint{1412.4708})

\bibitem{Bernhard:2016tnd}
Bernhard J~E, Moreland J~S, Bass S~A, Liu J and Heinz U 2016 {\em Phys. Rev.
  C\/} {\bf 94} 024907 (\textit{Preprint} \eprint{1605.03954})

\bibitem{Everett:2020yty}
Everett D {\em et~al.\/} (JETSCAPE) 2020  (\textit{Preprint}
  \eprint{2010.03928})

\bibitem{Everett:2020xug}
Everett D {\em et~al.\/} (JETSCAPE) 2020  (\textit{Preprint}
  \eprint{2011.01430})

\bibitem{Nijs:2020roc}
Nijs G, Van Der~Schee W, G\"ursoy U and Snellings R 2020  (\textit{Preprint}
  \eprint{2010.15134})

\bibitem{Moreland:2018gsh}
Moreland J~S, Bernhard J~E and Bass S~A 2020 {\em Phys. Rev. C\/} {\bf 101}
  024911 (\textit{Preprint} \eprint{1808.02106})

\bibitem{McLerran:1994ni}
McLerran L~D and Venugopalan R 1994 {\em Phys. Rev.\/} {\bf D49} 2233--2241

\bibitem{McLerran:1994ka}
McLerran L~D and Venugopalan R 1994 {\em Phys. Rev.\/} {\bf D49} 3352--3355

\bibitem{Kovner:1995ja}
Kovner A, McLerran L~D and Weigert H 1995 {\em Phys. Rev. D\/} {\bf 52}
  6231--6237 (\textit{Preprint} \eprint{hep-ph/9502289})

\bibitem{Iancu:2003xm}
Iancu E and Venugopalan R 2003 {\em {The Color glass condensate and high-energy
  scattering in QCD}\/} pp 249--3363 (\textit{Preprint}
  \eprint{hep-ph/0303204})

\bibitem{Schenke:2014zha}
Schenke B and Venugopalan R 2014 {\em Phys. Rev. Lett.\/} {\bf 113} 102301
  (\textit{Preprint} \eprint{1405.3605})

\bibitem{Romatschke:2013re}
Romatschke P and Hogg J~D 2013 {\em JHEP\/} {\bf 04} 048 (\textit{Preprint}
  \eprint{1301.2635})

\bibitem{Romatschke:2017ejr}
Romatschke P and Romatschke U 2019 {\em {Relativistic Fluid Dynamics In and Out
  of Equilibrium}\/} Cambridge Monographs on Mathematical Physics (Cambridge
  University Press) ISBN 978-1-108-48368-1, 978-1-108-75002-8
  (\textit{Preprint} \eprint{1712.05815})

\bibitem{Maldacena:1997re}
Maldacena J~M 1999 {\em Int. J. Theor. Phys.\/} {\bf 38} 1113--1133
  (\textit{Preprint} \eprint{hep-th/9711200})

\bibitem{Adler:2013aqf}
Adler S~S {\em et~al.\/} (PHENIX) 2014 {\em Phys. Rev. C\/} {\bf 89} 044905
  (\textit{Preprint} \eprint{1312.6676})

\bibitem{Eremin:2003qn}
Eremin S and Voloshin S 2003 {\em Phys. Rev. C\/} {\bf 67} 064905
  (\textit{Preprint} \eprint{nucl-th/0302071})

\bibitem{Bialas:1980zw}
Bialas A, Czyz W and Lesniak L 1982 {\em Phys. Rev. D\/} {\bf 25} 2328

\bibitem{Adamczyk:2015obl}
Adamczyk L {\em et~al.\/} (STAR) 2015 {\em Phys. Rev. Lett.\/} {\bf 115} 222301
  (\textit{Preprint} \eprint{1505.07812})

\bibitem{Albacete:2016pmp}
Albacete J~L and Soto-Ontoso A 2017 {\em Phys. Lett. B\/} {\bf 770} 149--153
  (\textit{Preprint} \eprint{1605.09176})

\bibitem{Arriola:2016bxa}
Ruiz~Arriola E and Broniowski W 2016 {\em Few Body Syst.\/} {\bf 57} 485--490
  (\textit{Preprint} \eprint{1602.00288})

\bibitem{Alkin:2014rfa}
Alkin A, Martynov E, Kovalenko O and Troshin S~M 2014 {\em Phys. Rev. D\/} {\bf
  89} 091501 (\textit{Preprint} \eprint{1403.8036})

\bibitem{Dremin:2015ujt}
Dremin I~M 2017 {\em Bull. Lebedev Phys. Inst.\/} {\bf 44} 94--98
  (\textit{Preprint} \eprint{1511.03212})

\bibitem{Troshin:2016frs}
Troshin S~M and Tyurin N~E 2016 {\em Mod. Phys. Lett. A\/} {\bf 31} 1650079
  (\textit{Preprint} \eprint{1602.08972})

\bibitem{Antchev:2011zz}
Antchev G {\em et~al.\/} (TOTEM) 2011 {\em EPL\/} {\bf 95} 41001
  (\textit{Preprint} \eprint{1110.1385})

\bibitem{McGlinchey:2016ssj}
McGlinchey D, Nagle J~L and Perepelitsa D~V 2016 {\em Phys. Rev. C\/} {\bf 94}
  024915 (\textit{Preprint} \eprint{1603.06607})

\bibitem{Mantysaari:2016ykx}
M\"antysaari H and Schenke B 2016 {\em Phys. Rev. Lett.\/} {\bf 117} 052301
  (\textit{Preprint} \eprint{1603.04349})

\bibitem{Miettinen:1978jb}
Miettinen H~I and Pumplin J 1978 {\em Phys. Rev. D\/} {\bf 18} 1696

\bibitem{Frankfurt:1993qi}
Frankfurt L, Miller G~A and Strikman M 1993 {\em Phys. Rev. Lett.\/} {\bf 71}
  2859--2862 (\textit{Preprint} \eprint{hep-ph/9309285})

\bibitem{Frankfurt:2008vi}
Frankfurt L, Strikman M, Treleani D and Weiss C 2008 {\em Phys. Rev. Lett.\/}
  {\bf 101} 202003 (\textit{Preprint} \eprint{0808.0182})

\bibitem{Caldwell:2009ke}
Caldwell A and Kowalski H 2009 {The J/psi Way to Nuclear Structure} {\em {13th
  International Conference on Elastic and Diffractive Scattering (Blois
  Workshop): Moving Forward into the LHC Era}\/} pp 190--192 (\textit{Preprint}
  \eprint{0909.1254})

\bibitem{Lappi:2010dd}
Lappi T and Mantysaari H 2011 {\em Phys. Rev. C\/} {\bf 83} 065202
  (\textit{Preprint} \eprint{1011.1988})

\bibitem{Chekanov:2002rm}
Chekanov S {\em et~al.\/} (ZEUS) 2003 {\em Eur. Phys. J. C\/} {\bf 26} 389--409
  (\textit{Preprint} \eprint{hep-ex/0205081})

\bibitem{Aktas:2003zi}
Aktas A {\em et~al.\/} (H1) 2003 {\em Phys. Lett. B\/} {\bf 568} 205--218
  (\textit{Preprint} \eprint{hep-ex/0306013})

\bibitem{Aktas:2005xu}
Aktas A {\em et~al.\/} (H1) 2006 {\em Eur. Phys. J. C\/} {\bf 46} 585--603
  (\textit{Preprint} \eprint{hep-ex/0510016})

\bibitem{Chekanov:2002xi}
Chekanov S {\em et~al.\/} (ZEUS) 2002 {\em Eur. Phys. J. C\/} {\bf 24} 345--360
  (\textit{Preprint} \eprint{hep-ex/0201043})

\bibitem{Alexa:2013xxa}
Alexa C {\em et~al.\/} (H1) 2013 {\em Eur. Phys. J. C\/} {\bf 73} 2466
  (\textit{Preprint} \eprint{1304.5162})

\bibitem{Mantysaari:2020axf}
M\"antysaari H 2020 {\em Rept. Prog. Phys.\/} {\bf 83} 082201
  (\textit{Preprint} \eprint{2001.10705})

\bibitem{Miller:2003sa}
Miller G~A 2003 {\em Phys. Rev. C\/} {\bf 68} 022201 (\textit{Preprint}
  \eprint{nucl-th/0304076})

\bibitem{Habich:2015rtj}
Habich M, Miller G~A, Romatschke P and Xiang W 2016 {\em Eur. Phys. J. C\/}
  {\bf 76} 408 (\textit{Preprint} \eprint{1512.05354})

\bibitem{Mantysaari:2016jaz}
M\"antysaari H and Schenke B 2016 {\em Phys. Rev. D\/} {\bf 94} 034042
  (\textit{Preprint} \eprint{1607.01711})

\bibitem{Mantysaari:2017cni}
M\"antysaari H, Schenke B, Shen C and Tribedy P 2017 {\em Phys. Lett. B\/} {\bf
  772} 681--686 (\textit{Preprint} \eprint{1705.03177})

\bibitem{Schenke:2020mbo}
Schenke B, Shen C and Tribedy P 2020 {\em Phys. Rev. C\/} {\bf 102} 044905
  (\textit{Preprint} \eprint{2005.14682})

\bibitem{Acharya:2019vdf}
Acharya S {\em et~al.\/} (ALICE) 2019 {\em Phys. Rev. Lett.\/} {\bf 123} 142301
  (\textit{Preprint} \eprint{1903.01790})

\bibitem{Mantysaari:2018zdd}
M\"antysaari H and Schenke B 2018 {\em Phys. Rev. D\/} {\bf 98} 034013
  (\textit{Preprint} \eprint{1806.06783})

\bibitem{Aschenauer:2017jsk}
Aschenauer E~C, Fazio S, Lee J~H, Mantysaari H, Page B~S, Schenke B, Ullrich T,
  Venugopalan R and Zurita P 2019 {\em Rept. Prog. Phys.\/} {\bf 82} 024301
  (\textit{Preprint} \eprint{1708.01527})

\bibitem{Mantysaari:2019jhh}
M\"antysaari H and Schenke B 2020 {\em Phys. Rev. C\/} {\bf 101} 015203
  (\textit{Preprint} \eprint{1910.03297})

\bibitem{Mantysaari:2017dwh}
M\"antysaari H and Schenke B 2017 {\em Phys. Lett. B\/} {\bf 772} 832--838
  (\textit{Preprint} \eprint{1703.09256})

\bibitem{Mantysaari:2020lhf}
M\"antysaari H, Roy K, Salazar F and Schenke B 2020  (\textit{Preprint}
  \eprint{2011.02464})

\bibitem{Jalilian-Marian:1997jx}
Jalilian-Marian J, Kovner A, Leonidov A and Weigert H 1997 {\em Nucl. Phys.\/}
  {\bf B504} 415--431 (\textit{Preprint} \eprint{hep-ph/9701284})

\bibitem{Jalilian-Marian:1997gr}
Jalilian-Marian J, Kovner A, Leonidov A and Weigert H 1999 {\em Phys. Rev.\/}
  {\bf D59} 014014 (\textit{Preprint} \eprint{hep-ph/9706377})

\bibitem{Iancu:2000hn}
Iancu E, Leonidov A and McLerran L~D 2001 {\em Nucl. Phys. A\/} {\bf 692}
  583--645 (\textit{Preprint} \eprint{hep-ph/0011241})

\bibitem{Mueller:2001uk}
Mueller A~H 2001 {\em Phys. Lett. B\/} {\bf 523} 243--248 (\textit{Preprint}
  \eprint{hep-ph/0110169})

\bibitem{Kowalski:2003hm}
Kowalski H and Teaney D 2003 {\em Phys. Rev. D\/} {\bf 68} 114005
  (\textit{Preprint} \eprint{hep-ph/0304189})

\bibitem{Rezaeian:2012ji}
Rezaeian A~H, Siddikov M, Van~de Klundert M and Venugopalan R 2013 {\em Phys.
  Rev. D\/} {\bf 87} 034002 (\textit{Preprint} \eprint{1212.2974})

\bibitem{Mantysaari:2018nng}
M\"antysaari H and Zurita P 2018 {\em Phys. Rev. D\/} {\bf 98} 036002
  (\textit{Preprint} \eprint{1804.05311})

\bibitem{Schlichting:2014ipa}
Schlichting S and Schenke B 2014 {\em Phys. Lett. B\/} {\bf 739} 313--319
  (\textit{Preprint} \eprint{1407.8458})

\bibitem{Froissart:1961ux}
Froissart M 1961 {\em Phys. Rev.\/} {\bf 123} 1053--1057

\bibitem{Martin:1962rt}
Martin A 1963 {\em Phys. Rev.\/} {\bf 129} 1432--1436

\bibitem{Gotsman:2020ryd}
Gotsman E and Levin E 2021 {\em Phys. Rev. D\/} {\bf 103} 014020
  (\textit{Preprint} \eprint{2009.12218})

\bibitem{Schenke:2016ksl}
Schenke B and Schlichting S 2016 {\em Phys. Rev. C\/} {\bf 94} 044907
  (\textit{Preprint} \eprint{1605.07158})

\bibitem{McDonald:2020oyf}
McDonald S, Jeon S and Gale C 2021 {\em Nucl. Phys. A\/} {\bf 1005} 121771
  (\textit{Preprint} \eprint{2001.08636})

\bibitem{Giacalone:2019kgg}
Giacalone G, Guerrero-Rodr\'\i{}guez P, Luzum M, Marquet C and Ollitrault J~Y
  2019 {\em Phys. Rev. C\/} {\bf 100} 024905 (\textit{Preprint}
  \eprint{1902.07168})

\bibitem{Gelis:2019vzt}
Gelis F, Giacalone G, Guerrero-Rodr\'\i{}guez P, Marquet C and Ollitrault J~Y
  2019  (\textit{Preprint} \eprint{1907.10948})

\bibitem{Albacete:2018bbv}
Albacete J~L, Guerrero-Rodr\'\i{}guez P and Marquet C 2019 {\em JHEP\/} {\bf
  01} 073 (\textit{Preprint} \eprint{1808.00795})

\bibitem{Giacalone:2021}
Giacalone G 2021
  \urlprefix\url{https://indico.cern.ch/event/854124/contributions/4136904/}

\bibitem{Nagle:2018ybc}
Nagle J~L and Zajc W~A 2019 {\em Phys. Rev. C\/} {\bf 99} 054908
  (\textit{Preprint} \eprint{1808.01276})

\bibitem{Diehl:1997bu}
Diehl M, Gousset T, Pire B and Ralston J~P 1997 {\em Phys. Lett. B\/} {\bf 411}
  193--202 (\textit{Preprint} \eprint{hep-ph/9706344})

\bibitem{Hoodbhoy:1998vm}
Hoodbhoy P and Ji X~D 1998 {\em Phys. Rev. D\/} {\bf 58} 054006
  (\textit{Preprint} \eprint{hep-ph/9801369})

\bibitem{Belitsky:2000jk}
Belitsky A~V and Mueller D 2000 {\em Phys. Lett. B\/} {\bf 486} 369--377
  (\textit{Preprint} \eprint{hep-ph/0005028})

\bibitem{Diehl:2001pm}
Diehl M 2001 {\em Eur. Phys. J. C\/} {\bf 19} 485--492 (\textit{Preprint}
  \eprint{hep-ph/0101335})

\bibitem{Belitsky:2001ns}
Belitsky A~V, Mueller D and Kirchner A 2002 {\em Nucl. Phys. B\/} {\bf 629}
  323--392 (\textit{Preprint} \eprint{hep-ph/0112108})

\bibitem{Meissner:2009ww}
Meissner S, Metz A and Schlegel M 2009 {\em JHEP\/} {\bf 08} 056
  (\textit{Preprint} \eprint{0906.5323})

\bibitem{Meissner:2008ay}
Meissner S, Metz A, Schlegel M and Goeke K 2008 {\em JHEP\/} {\bf 08} 038
  (\textit{Preprint} \eprint{0805.3165})

\bibitem{Lorce:2011dv}
Lorce C, Pasquini B and Vanderhaeghen M 2011 {\em JHEP\/} {\bf 05} 041
  (\textit{Preprint} \eprint{1102.4704})

\bibitem{Lorce:2013pza}
Lorc\'e C and Pasquini B 2013 {\em JHEP\/} {\bf 09} 138 (\textit{Preprint}
  \eprint{1307.4497})

\bibitem{Ji:2003ak}
Ji X~d 2003 {\em Phys. Rev. Lett.\/} {\bf 91} 062001 (\textit{Preprint}
  \eprint{hep-ph/0304037})

\bibitem{Belitsky:2003nz}
Belitsky A~V, Ji X~d and Yuan F 2004 {\em Phys. Rev. D\/} {\bf 69} 074014
  (\textit{Preprint} \eprint{hep-ph/0307383})

\bibitem{Krasnitz:2002ng}
Krasnitz A, Nara Y and Venugopalan R 2003 {\em Phys. Lett. B\/} {\bf 554}
  21--27 (\textit{Preprint} \eprint{hep-ph/0204361})

\bibitem{Gelis:2008ad}
Gelis F, Lappi T and Venugopalan R 2008 {\em Phys. Rev. D\/} {\bf 78} 054020
  (\textit{Preprint} \eprint{0807.1306})

\bibitem{Gelis:2008sz}
Gelis F, Lappi T and Venugopalan R 2009 {\em Phys. Rev. D\/} {\bf 79} 094017
  (\textit{Preprint} \eprint{0810.4829})

\bibitem{Dumitru:2008wn}
Dumitru A, Gelis F, McLerran L and Venugopalan R 2008 {\em Nucl. Phys. A\/}
  {\bf 810} 91--108 (\textit{Preprint} \eprint{0804.3858})

\bibitem{AD_rikenwkshp}
Dumitru A {in {RIKEN-BNL} Center Workshop on ``Progress in High pT Physics at
  RHIC'', March 17 -- 19, 2010, RBRC Vol. 95, page 129.}

\bibitem{Dumitru:2010iy}
Dumitru A, Dusling K, Gelis F, Jalilian-Marian J, Lappi T and Venugopalan R
  2011 {\em Phys. Lett. B\/} {\bf 697} 21--25 (\textit{Preprint}
  \eprint{1009.5295})

\bibitem{Levin:2011fb}
Levin E and Rezaeian A~H 2011 {\em Phys. Rev. D\/} {\bf 84} 034031
  (\textit{Preprint} \eprint{1105.3275})

\bibitem{Gyulassy:2014cfa}
Gyulassy M, Levai P, Vitev I and Biro T~S 2014 {\em Phys. Rev. D\/} {\bf 90}
  054025 (\textit{Preprint} \eprint{1405.7825})

\bibitem{Dumitru:2010mv}
Dumitru A and Jalilian-Marian J 2010 {\em Phys. Rev. D\/} {\bf 81} 094015
  (\textit{Preprint} \eprint{1001.4820})

\bibitem{Dusling:2012wy}
Dusling K and Venugopalan R 2013 {\em Phys. Rev. D\/} {\bf 87} 054014
  (\textit{Preprint} \eprint{1211.3701})

\bibitem{Dusling:2013qoz}
Dusling K and Venugopalan R 2013 {\em Phys. Rev. D\/} {\bf 87} 094034
  (\textit{Preprint} \eprint{1302.7018})

\bibitem{McLerran:1998nk}
McLerran L~D and Venugopalan R 1999 {\em Phys. Rev. D\/} {\bf 59} 094002
  (\textit{Preprint} \eprint{hep-ph/9809427})

\bibitem{Dominguez:2008aa}
Dominguez F, Marquet C and Wu B 2009 {\em Nucl. Phys. A\/} {\bf 823} 99--119
  (\textit{Preprint} \eprint{0812.3878})

\bibitem{Lappi:2015vta}
Lappi T, Schenke B, Schlichting S and Venugopalan R 2016 {\em JHEP\/} {\bf 01}
  061 (\textit{Preprint} \eprint{1509.03499})

\bibitem{Krasnitz:1998ns}
Krasnitz A and Venugopalan R 1999 {\em Nucl. Phys. B\/} {\bf 557} 237
  (\textit{Preprint} \eprint{hep-ph/9809433})

\bibitem{Krasnitz:2002mn}
Krasnitz A, Nara Y and Venugopalan R 2003 {\em Nucl. Phys. A\/} {\bf 717}
  268--290 (\textit{Preprint} \eprint{hep-ph/0209269})

\bibitem{Lappi:2003bi}
Lappi T 2003 {\em Phys. Rev. C\/} {\bf 67} 054903 (\textit{Preprint}
  \eprint{hep-ph/0303076})

\bibitem{Schenke:2015aqa}
Schenke B, Schlichting S and Venugopalan R 2015 {\em Phys. Lett. B\/} {\bf 747}
  76--82 (\textit{Preprint} \eprint{1502.01331})

\bibitem{Lappi:2015vha}
Lappi T 2015 {\em Phys. Lett. B\/} {\bf 744} 315--319 (\textit{Preprint}
  \eprint{1501.05505})

\bibitem{Lappi:2009xa}
Lappi T, Srednyak S and Venugopalan R 2010 {\em JHEP\/} {\bf 01} 066
  (\textit{Preprint} \eprint{0911.2068})

\bibitem{McLerran:2016snu}
McLerran L and Skokov V 2017 {\em Nucl. Phys. A\/} {\bf 959} 83--101
  (\textit{Preprint} \eprint{1611.09870})

\bibitem{Kovner:2016jfp}
Kovner A, Lublinsky M and Skokov V 2017 {\em Phys. Rev. D\/} {\bf 96} 016010
  (\textit{Preprint} \eprint{1612.07790})

\bibitem{Kovchegov:2018jun}
Kovchegov Y~V and Skokov V~V 2018 {\em Phys. Rev. D\/} {\bf 97} 094021
  (\textit{Preprint} \eprint{1802.08166})

\bibitem{Schlichting:2019bvy}
Schlichting S and Skokov V 2020 {\em Phys. Lett. B\/} {\bf 806} 135511
  (\textit{Preprint} \eprint{1910.12496})

\bibitem{Agostini:2019avp}
Agostini P, Altinoluk T and Armesto N 2019 {\em Eur. Phys. J. C\/} {\bf 79} 600
  (\textit{Preprint} \eprint{1902.04483})

\bibitem{Agostini:2019hkj}
Agostini P, Altinoluk T and Armesto N 2019 {\em Eur. Phys. J. C\/} {\bf 79} 790
  (\textit{Preprint} \eprint{1907.03668})

\bibitem{Gotsman:2016owk}
Gotsman E, Levin E and Maor U 2017 {\em Phys. Rev. D\/} {\bf 95} 034005
  (\textit{Preprint} \eprint{1604.04461})

\bibitem{Gotsman:2016fee}
Gotsman E, Levin E and Maor U 2016 {\em Eur. Phys. J. C\/} {\bf 76} 607
  (\textit{Preprint} \eprint{1607.00594})

\bibitem{Dusling:2017dqg}
Dusling K, Mace M and Venugopalan R 2018 {\em Phys. Rev. Lett.\/} {\bf 120}
  042002 (\textit{Preprint} \eprint{1705.00745})

\bibitem{Dusling:2017aot}
Dusling K, Mace M and Venugopalan R 2018 {\em Phys. Rev. D\/} {\bf 97} 016014
  (\textit{Preprint} \eprint{1706.06260})

\bibitem{Mace:2018vwq}
Mace M, Skokov V~V, Tribedy P and Venugopalan R 2018 {\em Phys. Rev. Lett.\/}
  {\bf 121} 052301 [Erratum: Phys.Rev.Lett. 123, 039901 (2019)]
  (\textit{Preprint} \eprint{1805.09342})

\bibitem{Iancu:2017fzn}
Iancu E and Rezaeian A~H 2017 {\em Phys. Rev. D\/} {\bf 95} 094003
  (\textit{Preprint} \eprint{1702.03943})

\bibitem{Altinoluk:2015uaa}
Altinoluk T, Armesto N, Beuf G, Kovner A and Lublinsky M 2015 {\em Phys. Lett.
  B\/} {\bf 751} 448--452 (\textit{Preprint} \eprint{1503.07126})

\bibitem{Kovchegov:2012nd}
Kovchegov Y~V and Wertepny D~E 2013 {\em Nucl. Phys. A\/} {\bf 906} 50--83
  (\textit{Preprint} \eprint{1212.1195})

\bibitem{HanburyBrown:1956bqd}
Hanbury~Brown R and Twiss R~Q 1956 {\em Nature\/} {\bf 178} 1046--1048

\bibitem{Dusling:2012cg}
Dusling K and Venugopalan R 2013 {\em Phys. Rev. D\/} {\bf 87} 051502
  (\textit{Preprint} \eprint{1210.3890})

\bibitem{Dusling:2013oia}
Dusling K and Venugopalan R 2013 {\em Phys. Rev. D\/} {\bf 87} 094034
  (\textit{Preprint} \eprint{1302.7018})

\bibitem{Greif:2020rhi}
Greif M, Greiner C, Pl\"atzer S, Schenke B and Schlichting S 2020
  (\textit{Preprint} \eprint{2012.08493})

\bibitem{Schenke:2016lrs}
Schenke B, Schlichting S, Tribedy P and Venugopalan R 2016 {\em Phys. Rev.
  Lett.\/} {\bf 117} 162301 (\textit{Preprint} \eprint{1607.02496})

\bibitem{PHENIX:2018lia}
Aidala C {\em et~al.\/} (PHENIX) 2019 {\em Nature Phys.\/} {\bf 15} 214--220
  (\textit{Preprint} \eprint{1805.02973})

\bibitem{Aad:2013fja}
Aad G {\em et~al.\/} (ATLAS) 2013 {\em Phys. Lett. B\/} {\bf 725} 60--78
  (\textit{Preprint} \eprint{1303.2084})

\bibitem{Chatrchyan:2013nka}
Chatrchyan S {\em et~al.\/} (CMS) 2013 {\em Phys. Lett. B\/} {\bf 724} 213--240
  (\textit{Preprint} \eprint{1305.0609})

\bibitem{Abelev:2014mda}
Abelev B~B {\em et~al.\/} (ALICE) 2014 {\em Phys. Rev. C\/} {\bf 90} 054901
  (\textit{Preprint} \eprint{1406.2474})

\bibitem{Khachatryan:2015waa}
Khachatryan V {\em et~al.\/} (CMS) 2015 {\em Phys. Rev. Lett.\/} {\bf 115}
  012301 (\textit{Preprint} \eprint{1502.05382})

\bibitem{Yan:2013laa}
Yan L and Ollitrault J~Y 2014 {\em Phys. Rev. Lett.\/} {\bf 112} 082301
  (\textit{Preprint} \eprint{1312.6555})

\bibitem{Skokov:2014tka}
Skokov V 2015 {\em Phys. Rev. D\/} {\bf 91} 054014 (\textit{Preprint}
  \eprint{1412.5191})

\bibitem{Dumitru:2014yza}
Dumitru A, McLerran L and Skokov V 2015 {\em Phys. Lett. B\/} {\bf 743}
  134--137 (\textit{Preprint} \eprint{1410.4844})

\bibitem{Bozek:2016yoj}
Bozek P 2016 {\em Phys. Rev. C\/} {\bf 93} 044908 (\textit{Preprint}
  \eprint{1601.04513})

\bibitem{Schukraft:2012ah}
Schukraft J, Timmins A and Voloshin S~A 2013 {\em Phys. Lett. B\/} {\bf 719}
  394--398 (\textit{Preprint} \eprint{1208.4563})

\bibitem{Schenke:2020uqq}
Schenke B, Shen C and Teaney D 2020 {\em Phys. Rev. C\/} {\bf 102} 034905
  (\textit{Preprint} \eprint{2004.00690})

\bibitem{Aad:2019fgl}
Aad G {\em et~al.\/} (ATLAS) 2019 {\em Eur. Phys. J. C\/} {\bf 79} 985
  (\textit{Preprint} \eprint{1907.05176})

\bibitem{Giacalone:2020dln}
Giacalone G, Gardim F~G, Noronha-Hostler J and Ollitrault J~Y 2020
  (\textit{Preprint} \eprint{2004.01765})

\bibitem{Giacalone:2020awm}
Giacalone G 2020 {\em Phys. Rev. C\/} {\bf 102} 024901 (\textit{Preprint}
  \eprint{2004.14463})

\bibitem{Giacalone:2020byk}
Giacalone G, Schenke B and Shen C 2020 {\em Phys. Rev. Lett.\/} {\bf 125}
  192301 (\textit{Preprint} \eprint{2006.15721})

\bibitem{Schenke:2010nt}
Schenke B, Jeon S and Gale C 2010 {\em Phys. Rev. C\/} {\bf 82} 014903
  (\textit{Preprint} \eprint{1004.1408})

\bibitem{Schenke:2010rr}
Schenke B, Jeon S and Gale C 2011 {\em Phys. Rev. Lett.\/} {\bf 106} 042301
  (\textit{Preprint} \eprint{1009.3244})

\bibitem{Schenke:2011bn}
Schenke B, Jeon S and Gale C 2012 {\em Phys. Rev. C\/} {\bf 85} 024901
  (\textit{Preprint} \eprint{1109.6289})

\bibitem{Bass:1998ca}
Bass S~A {\em et~al.\/} 1998 {\em Prog. Part. Nucl. Phys.\/} {\bf 41} 255--369
  (\textit{Preprint} \eprint{nucl-th/9803035})

\bibitem{Bleicher:1999xi}
Bleicher M {\em et~al.\/} 1999 {\em J. Phys. G\/} {\bf 25} 1859--1896
  (\textit{Preprint} \eprint{hep-ph/9909407})

\bibitem{Zhang:2021phk}
Zhang C, Behera A, Bhatta S and Jia J 2021  (\textit{Preprint}
  \eprint{2102.05200})

\bibitem{Sjostrand:2014zea}
Sj\"ostrand T, Ask S, Christiansen J~R, Corke R, Desai N, Ilten P, Mrenna S,
  Prestel S, Rasmussen C~O and Skands P~Z 2015 {\em Comput. Phys. Commun.\/}
  {\bf 191} 159--177 (\textit{Preprint} \eprint{1410.3012})

\bibitem{Bierlich:2016vgw}
Bierlich C, Gustafson G and L\"onnblad L 2016  (\textit{Preprint}
  \eprint{1612.05132})

\bibitem{Bierlich:2014xba}
Bierlich C, Gustafson G, L\"onnblad L and Tarasov A 2015 {\em JHEP\/} {\bf 03}
  148 (\textit{Preprint} \eprint{1412.6259})

\bibitem{Aidala:2017pup}
Aidala C {\em et~al.\/} (PHENIX) 2017 {\em Phys. Rev. C\/} {\bf 96} 064905
  (\textit{Preprint} \eprint{1708.06983})

\bibitem{Bozek:2014cya}
Bozek P and Broniowski W 2014 {\em Phys. Lett. B\/} {\bf 739} 308--312
  (\textit{Preprint} \eprint{1409.2160})

\bibitem{Bzdak:2014dia}
Bzdak A and Ma G~L 2014 {\em Phys. Rev. Lett.\/} {\bf 113} 252301
  (\textit{Preprint} \eprint{1406.2804})

\bibitem{Koop:2015wea}
Orjuela~Koop J~D, Adare A, McGlinchey D and Nagle J~L 2015 {\em Phys. Rev. C\/}
  {\bf 92} 054903 (\textit{Preprint} \eprint{1501.06880})

\bibitem{Shen:2017bsr}
Shen C and Schenke B 2018 {\em Phys. Rev. C\/} {\bf 97} 024907
  (\textit{Preprint} \eprint{1710.00881})

\bibitem{Schlichting:2020wrv}
Schlichting S and Singh P 2021 {\em Phys. Rev. D\/} {\bf 103} 014003
  (\textit{Preprint} \eprint{2010.11172})

\bibitem{Adler:2003ii}
Adler S~S {\em et~al.\/} (PHENIX) 2003 {\em Phys. Rev. Lett.\/} {\bf 91} 072303
  (\textit{Preprint} \eprint{nucl-ex/0306021})

\bibitem{Adams:2003im}
Adams J {\em et~al.\/} (STAR) 2003 {\em Phys. Rev. Lett.\/} {\bf 91} 072304
  (\textit{Preprint} \eprint{nucl-ex/0306024})

\bibitem{Arsene:2003yk}
Arsene I {\em et~al.\/} (BRAHMS) 2003 {\em Phys. Rev. Lett.\/} {\bf 91} 072305
  (\textit{Preprint} \eprint{nucl-ex/0307003})

\bibitem{Back:2003ns}
Back B~B {\em et~al.\/} (PHOBOS) 2003 {\em Phys. Rev. Lett.\/} {\bf 91} 072302
  (\textit{Preprint} \eprint{nucl-ex/0306025})

\bibitem{Khachatryan:2016odn}
Khachatryan V {\em et~al.\/} (CMS) 2017 {\em JHEP\/} {\bf 04} 039
  (\textit{Preprint} \eprint{1611.01664})

\bibitem{Tywoniuk:2014hta}
Tywoniuk K 2014 {\em Nucl. Phys. A\/} {\bf 926} 85--91

\bibitem{Shen:2016egw}
Shen C, Park C, Paquet J~F, Denicol G~S, Jeon S and Gale C 2016 {\em Nucl.
  Phys. A\/} {\bf 956} 741--744 (\textit{Preprint} \eprint{1601.03070})

\bibitem{ATLAS:2014cpa}
Aad G {\em et~al.\/} (ATLAS) 2015 {\em Phys. Lett. B\/} {\bf 748} 392--413
  (\textit{Preprint} \eprint{1412.4092})

\bibitem{Huang:2019tgz}
Huang S, Chen Z, Jia J and Li W 2020 {\em Phys. Rev. C\/} {\bf 101} 021901
  (\textit{Preprint} \eprint{1904.10415})

\bibitem{Shen:2015qba}
Shen C, Paquet J~F, Denicol G~S, Jeon S and Gale C 2016 {\em Phys. Rev.
  Lett.\/} {\bf 116} 072301 (\textit{Preprint} \eprint{1504.07989})

\bibitem{Adare:2012vn}
Adare A {\em et~al.\/} (PHENIX) 2013 {\em Phys. Rev. C\/} {\bf 87} 054907
  (\textit{Preprint} \eprint{1208.1234})

\end{thebibliography}
\end{document}